\documentclass[english,final]{IEEEtran}
\usepackage[T1]{fontenc}
\usepackage[utf8]{luainputenc}
\usepackage{float}
\usepackage{amsmath}
\usepackage{amssymb}
\usepackage{graphicx}
\usepackage{esint}
\usepackage{cite}
\usepackage{hyperref}

\makeatletter

\floatstyle{ruled}
\newfloat{algorithm}{tbp}{loa}
\providecommand{\algorithmname}{Algorithm}
\floatname{algorithm}{\protect\algorithmname}

\usepackage{babel}

\usepackage{color}

\@ifundefined{showcaptionsetup}{}{%
 \PassOptionsToPackage{caption=false}{subfig}}
\usepackage{subfig}
\makeatother

\usepackage{babel}
\begin{document}

\title{Manifold Learning with Contracting Observers for Data-driven Time-series
Analysis}

\author{Tal~Shnitzer, Ronen~Talmon,~\IEEEmembership{Member,~IEEE}, Jean-Jacques~Slotine
\thanks{Tal~Shnitzer and Ronen~Talmon are with the Department of Electrical Engineering, Technion -- Israel Institute of Technology, Technion City, Haifa, Israel 32000 (e-mail: shnitzer@campus.technion.ac.il; ronen@ee.technion.ac.il).}
\thanks{Jean-Jacques E. Slotine is with the Nonlinear Systems Laboratory, Massachusetts Institute of Technology, Cambridge, Massachusetts 02139, U.S.A. (e-mail: jjs@mit.edu).}
\thanks{This work was supported by the European Union's Seventh Framework Programme (FP7) under Marie Curie Grant 630657.}}

\maketitle
\begin{abstract}
Analyzing signals arising from dynamical systems typically requires many modeling assumptions and parameter estimation. In high dimensions, this modeling is particularly difficult due to the ``curse of dimensionality''. In this paper, we propose a method for building an intrinsic representation of such signals in a purely data-driven manner. First, we apply a manifold learning technique, diffusion maps, to learn the intrinsic model of the latent variables of the dynamical system, solely from the measurements. Second, we use concepts and tools from control theory and build a linear contracting observer to estimate the latent variables in a sequential manner from new incoming measurements. The effectiveness of the presented framework is demonstrated by applying it to a toy problem and to a music analysis application. In these examples we show that our method reveals the intrinsic variables of the analyzed dynamical systems.
\end{abstract}
\begin{IEEEkeywords}
Intrinsic modeling, manifold learning, linear observer.
\end{IEEEkeywords}

\section{Introduction}

High dimensional signals generated by dynamical systems arise in many fields of science. For example, biomedical signals such as EEG and EMG can be modeled by few latent processes measured by a large set of noisy sensors. In such applications the goal is to identify the latent intrinsic variables which describe the true, intrinsic state of the system.

Analyzing such signals typically requires vast modeling assumptions. For example, Bayesian filtering methods require a priori knowledge of the statistical model and often rely on parameter estimation \cite{Kalman1960,Doucet2000}. Finding
appropriate models and estimating their parameters from high dimensional
data is challenging, since the ``curse of dimensionality'' leads to failure of many data analysis techniques that perform well for low dimensional data.

We approach the problem of high dimensional signal analysis in dynamical systems from a geometric modeling standpoint, by applying manifold learning techniques. From this standpoint, the main assumption is that the accessible high dimensional data (the observations of the system) lie on an underlying nonlinear manifold of lower dimensions. 
In the past decade, various manifold learning methods
have been introduced, e.g., \cite{Tenenbaum2000,Belkin2003,Donoho2003,Coifman2006},
in which the geometry of the underlying manifold is captured in a
data-driven manner, and the data are embedded in a low dimensional
space, thereby attaining dimensionality reduction. In classical manifold learning,
time series are processed as data sets of samples, ignoring their
embodied dynamics and temporal order. Recently, several methods have addressed this problem and incorporated the time dependency of consecutive samples into the manifold learning framework \cite{TalmonPNAS,berry2015semiparametric,berry2015nonparametric,berry2014nonparametric,TalmonTechReport}.  For example, in \cite{berry2015semiparametric}, a semi-parametric framework, based on manifold learning is presented. In this work, Berry and Harlim
construct an ensemble Kalman filter which reveals the process of an
underlying stochastic parameter and its probability density by applying
diffusion maps \cite{Coifman2006} and projecting the filtering problem onto a basis created
by the diffusion maps coordinates. However, in their setting, the
measurement modality and the state model are assumed to be known up
to an unknown stochastic parameter. In another work \cite{TalmonPNAS}, a non-parametric, Bayesian framework is presented, which incorporates
the dependency of consecutive time samples into diffusion maps. However,
this framework assumes a Gaussian setting, i.e. given the underlying
state the measurements are assumed to be locally Gaussian, and it requires the estimation of
the mean value and the local covariance matrices of the observations. 

In this paper, we use geometric analysis tools to capture the inherent structure of the observations and their dynamics. We exploit the recovered geometry, along with the smoothness in time, to construct a representation of the underlying state. For this purpose, a filtering framework is introduced, based on diffusion maps \cite{Coifman2005,Coifman2006} in a setting which is non-parametric and non-Gaussian. We particularly address time series analysis and propose an approach consisting of two steps: (i) learning the intrinsic model of the latent variables of the signal solely from measurements, and then, (ii) estimating the latent variables in a sequential manner from new incoming measurements. Such an approach allows us to analyze and process real signals without existing adequate models in the current literature. 
More specifically, we show that our method reveals the intrinsic state, its dynamics and its relationship to the observations based on the measurements. 
Furthermore, we show that the dynamics of the constructed diffusion maps coordinates are {\em linear}, even when modeling {\em highly nonlinear} systems. 
This allows us to devise a framework, which incorporates a standard processing technique for time series -- an observer, and devise a method, which is especially designed to handle noisy data. 
Finally, we apply our framework to a toy example and to a practical application of music analysis, in which we show that our method reveals intrinsic variables describing dominant musical notes as well as different musical instruments.

The paper is organized as follows. In Section \ref{sec:Problem-Formulation} the general setting and formulation of the problem are presented. In Section \ref{sec:Recovery-of-System} the diffusion maps framework is introduced as a method to recover the dimension and dynamics of the system. Section \ref{sec:The-Observer} presents a contracting observer, which is constructed based on the recovered information and learned dynamics, in order to reconstruct the state of the dynamical system. Section \ref{sec:Experimental-Results} illustrates the advantages of the constructed observer by applying our framework to a toy example and to a music analysis application. Section \ref{sec:Conclusions} offers brief concluding remarks.

\section{Problem Formulation\label{sec:Problem-Formulation}}

Many problems involving high dimensional signals can be modeled using
a state-space formulation. In this framework we are given a set of
high-dimensional measurements $\boldsymbol{z}\left(t\right)\in\mathbb{R}^{n}$.
We assume that the measurements are samples from a dynamical system
of the form 
\begin{eqnarray}
\dot{\boldsymbol{\theta}}(t) & = & f\left(\boldsymbol{\theta}\left(t\right),\dot{\boldsymbol{\omega}}(t)\right)\label{eq:dyn_sys1}\\
\boldsymbol{z}(t) & = & g\left(\boldsymbol{\theta}\left(t\right),\boldsymbol{v}\left(t\right)\right)\label{eq:dyn_sys2}
\end{eqnarray}
where $\boldsymbol{v}\in\mathbb{R}^{n}$ is a white noise process, $\boldsymbol{\omega}\in\mathbb{R}^{d}$ is Brownian motion and $\dot{\boldsymbol{\omega}}$ is the time derivative, $\boldsymbol{\theta}(t)\in\mathbb{R}^{d}$ is the true state of the system and $f,g$ are non-linear functions which represent the system dynamics and measurement function respectively.

We focus on the case in which $f$ is modeled as the following SDE:
\begin{align}
\dot{\boldsymbol{\theta}}(t) & =a\left(\boldsymbol{\theta}\left(t\right)\right)+b\left(\boldsymbol{\theta}\left(t\right)\right)\dot{\boldsymbol{\omega}}(t)\label{eq:StochEq}\\
\boldsymbol{z}(t) & =h(\boldsymbol{\theta}(t))+\boldsymbol{v}\left(t\right)\label{eq:MeasurementEq}
\end{align}
where $a\left(\boldsymbol{\theta}\right)$ and $b\left(\boldsymbol{\theta}\right)$ are the drift and diffusion coefficients. In this particular case, $a\left(\boldsymbol{\theta}\right)=-\nabla U\left(\boldsymbol{\theta}\right)$
where $U\left(\boldsymbol{\theta}\right)$ is referred to as a potential field, and $b\left(\boldsymbol{\theta}\right)=\sqrt{\frac{2}{\beta}}$ is a constant diffusion coefficient. 
The SDE in \eqref{eq:StochEq} is known as the Langevin equation, which 
describes the evolution of the state $\boldsymbol{\theta}\left(t\right)$
according to the potential $U\left(\boldsymbol{\theta}\right)$ at
an inverse temperature $\beta$. This terminology is derived from
the physical problem of particle motion in fluid, where the potential
$U\left(\boldsymbol{\theta}\right)$ represents the motion component which drives the particles to high density areas.
We assume that the potential $U\left(\boldsymbol{\theta}\right)$
is smooth and bounded (relaxation of this assumption is described
in \cite{coifman2008diffusion}), consequently, it is reasonable to
assume that \eqref{eq:StochEq} describes a stochastic diffusion process,
confined to a finite, compact, connected region $\mathcal{M}\subseteq\mathbb{R}^{d}$
with smooth reflecting boundaries \cite{coifman2008diffusion}.

Our goal here is to reveal the underlying dynamical process, $\boldsymbol{\theta}\left(t\right)$,
given the noisy measurements, $\boldsymbol{z}\left(t\right)$, in
a data-driven setting without additional model assumptions. In this
setting, all the parameters of the dynamical system are unknown, i.e.,
the potential function $U\left(\boldsymbol{\theta}\right)$, the thermal
factor $\beta$, the measurement function $h:\mathbb{\,R}^{d}\rightarrow\mathbb{R}^{n}$,
as well as the dimensionality $d$ and the coordinate system of the
state $\boldsymbol{\theta}(t)$.

We present the Ornstein Uhlenbeck equation, a simple case of \eqref{eq:StochEq},
as an example, which is implemented in Section \ref{sec:Experimental-Results}
to demonstrate the properties and efficacy of our proposed method.
The Ornstein Uhlenbeck equation is described as follows:
\[
\dot{\boldsymbol{\theta}}(t)=k\left(\mu-\boldsymbol{\theta}\left(t\right)\right)+\sigma\dot{\boldsymbol{\omega}}(t)
\]
where $\mu$ is the long term mean of the process, $k>0$ is the rate
that the process reverts to its mean value, and $\sigma$ is the diffusion
coefficient. This equation describes a noisy relaxation process, e.g.
over damped spring in the presence of thermal fluctuations.

We note that for simplicity, we omit the notation of $\left(t\right)$ in the following sections. However, all presented coordinates and processes are still time dependent.

The model described in \eqref{eq:StochEq} can represent a vast group
of physical phenomena, e.g. thermal noise processes, non-ideal harmonic
oscillators, diffusive particle motions, and financial processes \cite{Coffey2012langevin}.
In addition, it was previously used for empirical modeling of EEG
signals \cite{talmonDynSys2015} and of audio signals \cite{TalmonMagazine,TalmonReparam}. Due to its generality, most dynamical systems
can be roughly described by this model, however, our proposed method
is not confined to the model and we present results on a real application
in which the model can only be assumed. We note that, here, the main
purpose of the Langevin equation model \eqref{eq:StochEq} is to provide
a solid theoretical foundation.

\section{Discovering the System Model with Diffusion Maps\label{sec:Recovery-of-System}}

In this section we present the basic analysis showing how diffusion
maps can be used to reveal the model of a dynamical system from measurements
in a data-driven manner. This analysis is carried out assuming that
there is no measurement noise. When noise is present, the following
analysis no longer holds and the ability of the diffusion maps to
accurately recover the model parameters is hampered. In Section \ref{sec:The-Observer},
we extend this framework for noisy systems by explicitly incorporating
the system dynamics through the implementation of a contracting observer.

\subsection{Revealing the Dynamics of the Intrinsic State\label{sub:Revealing-the-dynamics}}

Consider the state dynamics \eqref{eq:StochEq}. Under the assumptions outlined
in Section \ref{sec:Problem-Formulation}, the local equilibrium transition
probability of this process is given by $p_{eq}\left(\boldsymbol{\theta}\right)=e^{-U\left(\boldsymbol{\theta}\right)}$.
For such a differential equation, the transition probability density
$p\left(\boldsymbol{\theta},t|\boldsymbol{\theta}_{0},t_{0}\right)$
of finding the system at time $t$ and at location $\boldsymbol{\theta}$,
given an initial location $\boldsymbol{\theta}_{0}$ at time $t_{0}$,
satisfies the backward Fokker-Planck equation:

\begin{equation}
\frac{\partial p}{\partial t}=\mathcal{L}p=\frac{1}{\beta}\triangle p-\nabla U\cdot \nabla p\label{eq:Fokker_Planck}
\end{equation}

Since we assume that the potential $U$ is smooth, it can be shown
\cite{Nadler2005,Nadler2006} that the operator $\mathcal{L}$ has
a discrete spectrum of non-positive decreasing eigenvalues $\left\{ -\lambda_{\ell}\right\} _{\ell=0}^{\infty}$
with associated eigenfunctions which satisfy: 
\begin{equation}
\mathcal{L}\psi_{\ell}=-\lambda_{\ell}\psi_{\ell}
\end{equation}
Based on It\^o's lemma, each of these eigenfunctions also evolve according
to the following stochastic differential equation \cite{coifman2008diffusion}:
\begin{equation}
\dot{\psi}_{\ell}=-\lambda_{\ell}\psi_{\ell}+\tilde{b}_{\ell}\left(\psi_{\ell},t\right)\dot{\omega}_{\ell},\,\,\ell=0,1,2\dots\label{eq:eigv_dyn}
\end{equation}
where $\tilde{b}_{\ell}(\psi_{\ell},t)$ has a known closed-form expression,
and $-\lambda_{\ell}$ are the eigenvalues of $\mathcal{L}$.
See details in Appendix \ref{sub:Appendix Eigenfunction-Dynamics}.

The eigenvalues and eigenfunctions of $\mathcal{L}$ along with their dynamics \eqref{eq:eigv_dyn} play a pivotal role in this work; their importance is three-fold. 
First, the eigenfunctions
of the Fokker-Planck operator form a parametrization of the
state, since the solution of \eqref{eq:Fokker_Planck} can be written
based on these eigenfunctions as $p\left(\boldsymbol{\theta},t|\boldsymbol{\theta}_{0},0\right)=\sum_{\ell=0}^{\infty}c_{\ell}e^{-\lambda_{\ell}t}\psi_{\ell}\left(\boldsymbol{\theta}\right)$,
where the coefficients $c_{\ell}$ are determined by the initial conditions
at $t=0$ \cite{coifman2008diffusion}. Second, the dynamics of this
parametrization is revealed in \eqref{eq:eigv_dyn}, which illustrates
that the resulting eigenfunctions, describing the long term behavior
of the given diffusion process, evolve according to a {\em linear drift,
determined by the eigenvalues of the operator}, with some additional
noise process. Third, in Section \ref{sub:Manifold-Learning-and},
we show that both the parametrization and the dynamics, which are
based on the eigenfunctions and eigenvalues of the Fokker-Planck operator,
can be approximated from the data without prior knowledge of the system.

\subsection{Data-driven Manifold Learning and Diffusion Maps
\label{sub:Manifold-Learning-and}}

Recall that the objective is to recover the underlying state given
the measurement process $\boldsymbol{z}\left(t\right)\in\mathbb{R}^{n}$.
For this purpose, as described in Section \ref{sub:Revealing-the-dynamics},
we wish to compute the eigenfunctions of the backward Fokker-Planck
operator describing the diffusion process, which are used to represent
the underlying state. In this section, we present the diffusion maps
framework \cite{Coifman2005,Coifman2006}, a computational
method to obtain these eigenfunctions from the data $\boldsymbol{\theta}\left(t\right)$, without prior information on the components comprising the right-hand side of \eqref{eq:Fokker_Planck}.

Based on the underlying state $\boldsymbol{\theta}\left(t\right)$, we build a pairwise affinity kernel $k_{\epsilon}\left(t,s\right)$ according to 
\begin{equation}
k_{\epsilon}\left(t,s\right)=\exp\left\{ -\frac{\left\Vert\boldsymbol{\theta}\left(t\right)-\boldsymbol{\theta}\left(s\right)\right\Vert^2}{\epsilon}\right\} \label{eq:PairwiseAffinity}
\end{equation}
where $\epsilon>0$. Here, $\left\Vert\cdot\right\Vert^2$ denotes the squared Euclidean norm, and $\epsilon$ is the kernel scale which
denotes a characteristic distance within the data set. In other words,
$\epsilon$ induces a notion of locality: if $\left\Vert\boldsymbol{\theta}\left(t\right)-\boldsymbol{\theta}\left(s\right)\right\Vert^2\gg\epsilon$,
then $k_\epsilon \left(t,s\right)$ is negligible.

The kernel is normalized as follows: 
\begin{equation}
p_{\epsilon}\left(t,s\right)=\frac{k_{\epsilon}\left(t,s\right)}{d_{\epsilon}\left(t\right)}\label{eq:Norm_all}
\end{equation}
where 
\[
d_{\epsilon}\left(t\right)={\int}k_{\epsilon}\left(t,s\right)q\left(s\right)ds
\]
and $q\left(s\right)=e^{-U\left(\boldsymbol{\theta}\left(s\right)\right)}$
is the equilibrium density of the underlying state parameter $\boldsymbol{\theta}\left(s\right)$.

Define the operator $P_{\epsilon}$ by 
\begin{equation}
\left(P_{\epsilon}g \right) \left(t\right)={\int}p_{\epsilon}\left(t,s\right)g\left(s\right)q\left(s\right)ds
\end{equation}

In the limit $\epsilon\rightarrow0$, the eigenfunctions of the operator 
\begin{equation}
L_{\epsilon}=\frac{1}{\epsilon}(P_{\epsilon}-\mathrm{I})\label{eq:Discrete_FP_Operator}
\end{equation}
converge to the eigenfunctions of the backward Fokker-Planck operator
\eqref{eq:Fokker_Planck} \cite{coifman2008diffusion,Singer2008},
where $\mathrm{I}$ is the identity operator.

The eigenvalue decomposition of the Fokker-Planck operator generates a discrete spectrum of eigenvalues $\left\{ -\lambda_{\ell}\right\} _{\ell=0}^{\infty}$  containing several dominant eigenvalues. In addition, these eigenvalues are decreasing,  therefore, we can assume that the first $m$ eigenvalues are the dominant ones and approximate the dynamical process \eqref{eq:StochEq} by a finite set of eigenfunctions and eigenvalues $\ell\in\left\{ 0,1,...,m\right\} $.
We construct a representation of the state based on these eigenfunctions, creating an embedded space, i.e. a new coordinate system:
\begin{equation}
[\psi_{1}(t),\psi_{2}(t),\dots,\psi_{m}(t)]^{T}\label{eq:mapping}
\end{equation}

In addition, as presented in \eqref{eq:eigv_dyn}, the dynamics of
these constructed coordinates can be approximated based on the corresponding
eigenvalues $\left\{ -\lambda_{\ell}\right\} _{\ell=0}^{m}$.

\subsection{Non-linear Measurement Mapping\label{sub:Non-linear-Measurement-mapping}}

In Section \ref{sub:Manifold-Learning-and} we show that the eigenfunctions of the backward Fokker-Planck operator give rise to a new coordinate system, with linear dynamics given by the eigenvalues, which appropriately describes the latent intrinsic state of the observed system. Since the underlying state $\boldsymbol{\theta}\left(t\right)$ is inaccessible and we are only given a set of measurements $\boldsymbol{z}(t)=h(\boldsymbol{\theta}\left(t\right))$, to construct the affinity kernel \eqref{eq:PairwiseAffinity} we approximate the required Euclidean distances of $\boldsymbol{\theta}\left(t\right)$ from the measurements by applying a modified version of the Mahalanobis distance presented in \cite{Singer2008}.

The modified Mahalanobis distance between two measurements, $\boldsymbol{z}\left(t\right)$
and $\boldsymbol{z}\left(s\right)$,
is given by:
\begin{align}
 & d\left(\boldsymbol{z}\left(t\right),\boldsymbol{z}\left(s\right)\right)\nonumber \\
 & =\frac{1}{2}\left(\boldsymbol{z}\left(t\right)-\boldsymbol{z}\left(s\right)\right)\left(C^{-1}\left(t\right)+C^{-1}\left(s\right)\right)\left(\boldsymbol{z}\left(t\right)-\boldsymbol{z}\left(s\right)\right)^{T}\label{eq:Mahalanibis Eq}
\end{align}
where $C\left(t\right)$ is the covariance matrix of the measurements $\boldsymbol{z}$ at time $t$ and $C\left(s\right)$ is the covariance at time $s$. We note that when $n$, the dimensionality
of the measurements $\boldsymbol{z}\left(t\right)$, is larger than
the state dimensionality $d$, the covariance matrices are not of
full rank and pseudo-inverse is used in \eqref{eq:Mahalanibis Eq}.
Singer et. al. show that this form of the Mahalanobis distance approximates
the Euclidean distance between two corresponding samples of the underlying
process, $\boldsymbol{\theta}\left(t\right)$: 
\[
d\left(\boldsymbol{z}\left(t\right),\boldsymbol{z}\left(s\right)\right)=\left\Vert \boldsymbol{\theta}\left(t\right)-\boldsymbol{\theta}\left(s\right)\right\Vert ^{2}+O\left(\left\Vert \boldsymbol{\theta}\left(t\right)-\boldsymbol{\theta}\left(s\right)\right\Vert ^{4}\right)
\]
The derivation of this modified Mahalanobis distance is presented
in Appendix \ref{sub:Appendix_Mahalanobis-Distance}. This holds assuming
that the Brownian motions of different coordinates, $\theta_{i},\,\,i=1,...,d$,
in \eqref{eq:StochEq} are independent. Therefore, by constructing
the diffusion maps based on this distance, we can recover the parameters
of the underlying state, i.e., its dynamics and diffusion maps coordinate
system, instead of those describing the measurements.

We can now construct a mapping from the measurements, $\boldsymbol{z}$, to the embedded space, based on the representation in \eqref{eq:mapping}:
\begin{equation}
\boldsymbol{z}(t)\mapsto[\psi_{1}(t),\psi_{2}(t),\dots,\psi_{m}(t)]^{T}\label{eq:mapping2}
\end{equation}

The above setting describes a purely data-driven scheme which provides
a representation of the underlying state based solely on the given
measurements. However, this framework has two main shortcomings. First,
system dynamics are not explicitly expressed in this mapping. Second,
the convergence of the constructed operator to the backward Fokker-Plank
operator is attained only when noiseless measurements are available.
We address these weaknesses in our observer setting presented in Section
\ref{sec:The-Observer}.

\subsection{Modeling the Lift Function\label{sub:Modeling-the-Lift}}

In the presented setting in Section \ref{sec:Problem-Formulation}, the dynamical system is modelled by \eqref{eq:StochEq} and \eqref{eq:MeasurementEq}. The state and its parameters in \eqref{eq:StochEq} are unknown and we have access only to the measurements $\boldsymbol{z}\left(t\right)$. By applying the modified Mahalanobis distance \eqref{eq:Mahalanibis Eq} we gain access to the Euclidean distances between the samples of the underlying state, from the given measurements. Based on these approximated Euclidean distances we can recover the parametrization and dynamics of the underlying state by applying diffusion maps as described in Section \ref{sub:Manifold-Learning-and}.
At this point, for the construction of a complete representation of the system, we are still missing \eqref{eq:MeasurementEq}, even in the noiseless case. In \eqref{eq:MeasurementEq}, $h$ is a function mapping the latent intrinsic state $\boldsymbol{\theta}\left(t\right)$ into the domain of the measurements $\boldsymbol{z}\left(t\right)$. Let $g\left(\cdot\right)$ be the lift function which is analogous to $h$, mapping the new recovered coordinate system to the domain of the measurements.

In this section we propose such a lift
function, based on the above eigenfunctions. Since the Fokker-Planck operator of the Langevin equation \eqref{eq:StochEq}
is Hermitian, its eigenfunctions form a basis for all real functions
defined on the underlying diffusion process \cite{Coifman2006a}.
Based on this concept we define a linear reconstruction function between
the new coordinate system and the measurements.

We expand each coordinate of the measurements, $z_{j}\left(t\right)\,\,j=1,...,n$,
using the eigenfunction basis,
\begin{equation}
z_{j}(t)=\sum_{\ell=1}^{\infty}\alpha_{j,\ell}\psi_{\ell}(t),\,\,j=1,\ldots,n
\end{equation}
where the expansion coefficients $\alpha_{j,\ell}$ are given by 
\begin{equation}
\alpha_{j,\ell}=\langle z_{j},\psi_{\ell}\rangle_{q}=\int_{-\infty}^{\infty}z_{j}(t)\psi_{\ell}(t)q\left(t\right)dt
\end{equation}
and $q\left(t\right)$ is the equilibrium density of the underlying
state parameter $\boldsymbol{\theta}\left(t\right)$.

By assuming that the spectrum of $L_{\epsilon}$ in \eqref{eq:Discrete_FP_Operator}
decays fast, and that for a finite $m$, the $m$ eigenfunctions associated
with the $m$ largest eigenvalues capture most of its energy, $z_{j}$
can be well approximated by 
\begin{equation}
z_{j}(t)\simeq\sum_{\ell=1}^{m}\alpha_{j,\ell}\psi_{\ell}(t)\label{eq:approx_expan_g}
\end{equation}

Let $\Psi(t)$ denote the parametrization of the state at time $t$
consisting these $m$ eigenfunctions,
\[
\Psi(t)=[\psi_{1}(t),\psi_{2}(t),\ldots,\psi_{m}(t)]^{T}
\]
We define $g(\cdot)$ as the mapping from this new coordinate system
to the measurements
\begin{equation}
\boldsymbol{z}(t)=g(\Psi(t)).
\end{equation}

Therefore, in matrix form, \eqref{eq:approx_expan_g} can be rewritten
as 
\begin{equation}
\boldsymbol{z}(t)=g(\Psi(t))\simeq\boldsymbol{\alpha}\Psi(t)\label{eq:approx_g}
\end{equation}
where $\boldsymbol{\alpha}$ is an $n\times m$ matrix whose elements
are given by $(\boldsymbol{\alpha})_{j,\ell}=\alpha_{j,\ell}$.

Choosing the optimal dimensionality $m$ is a widely studied problem \cite{hein2005intrinsic,coifman2008graph}. Here we apply a
heuristic approach, as described in \cite{coifman2008diffusion},
by determining the dimensionality based on the existence and location
of a spectral gap in the eigenvalues. This is shown to attain a meaningful,
low dimensional parametrization.

To conclude this section, we emphasize that the presented approach yields a data-driven coordinate system of the latent intrinsic state that does not require prior knowledge of the system.
The main benefit of this approach is that the constructed coordinate
system comprises both linear dynamics as well as a linear lift function.
Therefore, it enables us to provide linear solutions, as presented
in Section \ref{sec:The-Observer}, to highly nonlinear problems. We revisit this setting in Section \ref{sub:Koopman}, where we describe the relation
of our approach to the Koopman operator.

\section{The Observer Framework\label{sec:The-Observer}}

The derivation presented in Section \ref{sec:Recovery-of-System}
is based on a setup, which does not include measurement noise. When
considering real, noisy systems, the recovered parametrization is
merely an approximation of the results above, leading to inaccurate
representations of the underlying system state. In this section, we
propose a framework that improves the constructed parametrization
for noisy systems by facilitating the dynamics embodied in the measurements,
which did not take part in the construction of the embedding. For
this purpose, we use concepts and tools from control theory and propose
to build a linear contracting observer \cite{lohmiller1998contraction}.

Implementing an observer requires the model parameters, i.e. the dynamics
and lift function, and a parametrization of the underlying state.
Since we assume that these components are unknown we approximate them
using the manifold learning approach as described in Section \ref{sec:Recovery-of-System}.
In particular, the drift component describing system dynamics is shown
to be linear in \eqref{eq:eigv_dyn}, and therefore a linear observer
can be constructed.

\subsection{Contracting Observer\label{sub:Fixed-Contractive-Observer}}

Let $\widehat{\Psi}(t)\in\mathbb{R}^{m}$ denote the estimated state, which
is built based on the following standard recursive observer equation
\begin{equation}
\dot{\widehat{\Psi}}=\Lambda\widehat{\Psi}+\boldsymbol{\kappa}(\boldsymbol{z}-\hat{\boldsymbol{z}}).\label{eq:observer_rec}
\end{equation}
The components comprising the observer equation are as follows. First,
$\Lambda\in\mathbb{R}^{m\times m}$ is a diagonal matrix with the
$m$ largest eigenvalues of the Fokker-Planck operator on its diagonal:
$\Lambda_{\ell\ell}=-\lambda_{\ell}$, $\ell=1,\ldots,m$, since,
as discussed in Section \ref{sub:Revealing-the-dynamics}, they approximate
the dynamics of the system. Second, $\hat{\boldsymbol{z}}=g(\hat{\Psi})$
is the measurement associated with the current value of the estimated state,
where $g$ is the lift function defined in \eqref{eq:approx_g}. Finally,
$\boldsymbol{\kappa}\in\mathbb{R}^{m\times n}$ is an ``adjustable\textquotedbl{}
gain.

Note that the constructed coordinates, $\hat{\Psi}$, defined by the observer equation, are
neither orthogonal nor a basis, however, we assume that that the extension
of the function $g\left(\cdot\right)$ to $\hat{\Psi}$ approximates
the lift function from these coordinates to the measurements.
A related extension is presented in \cite{Coifman2006a}, which describes
a scheme for the extension of functions, defined on a given set, to additional elements.

By substituting \eqref{eq:approx_g} into \eqref{eq:observer_rec}, the
observer becomes :
\begin{align}
\dot{\widehat{\Psi}} & =\Lambda\widehat{\Psi}+\boldsymbol{\kappa}(\boldsymbol{z}-\boldsymbol{\alpha}\widehat{\Psi}) 
 =(\Lambda-\boldsymbol{\kappa}\boldsymbol{\alpha})\widehat{\Psi}+\boldsymbol{\kappa}\boldsymbol{z}
\end{align}
To obtain a contracting observer, we set the Jacobian $\mathbf{J}=(\Lambda-\boldsymbol{\kappa}\boldsymbol{\alpha})$
of the system to be negative. Specifically, by setting 
\begin{equation}
\boldsymbol{\kappa}=\gamma\Lambda\boldsymbol{\alpha}^{\dagger}\label{eq:kappa}
\end{equation}
where $\boldsymbol{\alpha}^{\dagger}$ denotes pseudo-inverse, we
have 
\begin{equation}
\mathbf{J}=(\Lambda-\boldsymbol{\kappa}\boldsymbol{\alpha})=(\Lambda-\gamma\Lambda\boldsymbol{\alpha}^{\dagger}\boldsymbol{\alpha})=(1-\gamma)\Lambda
\end{equation}
which guarantees contraction for any $\gamma<1$, since $\Lambda$
is negative. We remark that by setting this particular gain $\boldsymbol{\kappa}$
\eqref{eq:kappa}, we take advantage of the fact that in our particular
setup arising from the manifold learning standpoint, and in contrast
to the common practice in dynamical systems, the dimension $m$ of
the state (in our case, the inferred state $\hat{\Psi}$) is assumed
to be significantly smaller than the dimension $n$ of the measurement.
This implies the existence of a left pseudo-inverse satisfying $\boldsymbol{\alpha}^{\dagger}\boldsymbol{\alpha}=\mathbf{I}$.
By using such a constant value of $\boldsymbol{\kappa}$, the recursive
equation of the observer \eqref{eq:observer_rec} becomes 
\begin{align}
\dot{\widehat{\Psi}} & =\Lambda\widehat{\Psi}+\gamma\Lambda\boldsymbol{\alpha}^{\dagger}(\boldsymbol{z}-\boldsymbol{\alpha}\widehat{\Psi})=(1-\gamma)\Lambda\widehat{\Psi}+\gamma\Lambda\boldsymbol{\alpha}^{\dagger}\boldsymbol{z}\label{eq:Observer_actual}
\end{align}
where the value of $\gamma$ enables us to control the relative weighting
between the revealed dynamics ($\Lambda\hat{\Psi}$ in the first term
in the right-hand side of \eqref{eq:Observer_actual}), and the correspondence
to the measurements $\boldsymbol{z}$ (second term in the right-hand
side of \eqref{eq:Observer_actual}).
We note that the above form of $\boldsymbol{\kappa}$ is not optimal
and is chosen mainly for inversion of the lift (``measurement'') function. The
optimal choice of $\boldsymbol{\kappa}$ will be addressed in future
work.

\subsection{The Observer as a Koopman Operator\label{sub:Koopman}}
Based on the coordinate dynamics described in Section \ref{sec:Recovery-of-System},
our approach is related to Koopman spectral analysis \cite{Williams2015}.
Applying diffusion maps to the given measurements generates a parametrization
of the state space (intrinsic embedded coordinates) which evolves
according to known dynamics. These dynamics can be approximated by
a linear operator $\Lambda$ as described in Section \eqref{sub:Fixed-Contractive-Observer}.
In addition, in the presented linear case the mapping between the
embedded coordinates and the measurements, denoted by the function
$g$, is also linear.

Therefore, similarly to the Koopman operator, we create a linear operator
by applying $\Lambda$ to the observer's estimated state, describing the evolution of
the constructed coordinates on the state space of the dynamical system. However,
in contrast to common methods, which approximate the required quantities
for the construction of the Koopman operator, in our approach no additional
information is required and the constructed coordinates are purely
data driven. For example, the extended dynamic mode decomposition
(EDMD) \cite{Williams2015} requires additional information in the
form of dictionary elements.

\begin{figure}[t]
\centering{}\includegraphics[width=0.5\textwidth]{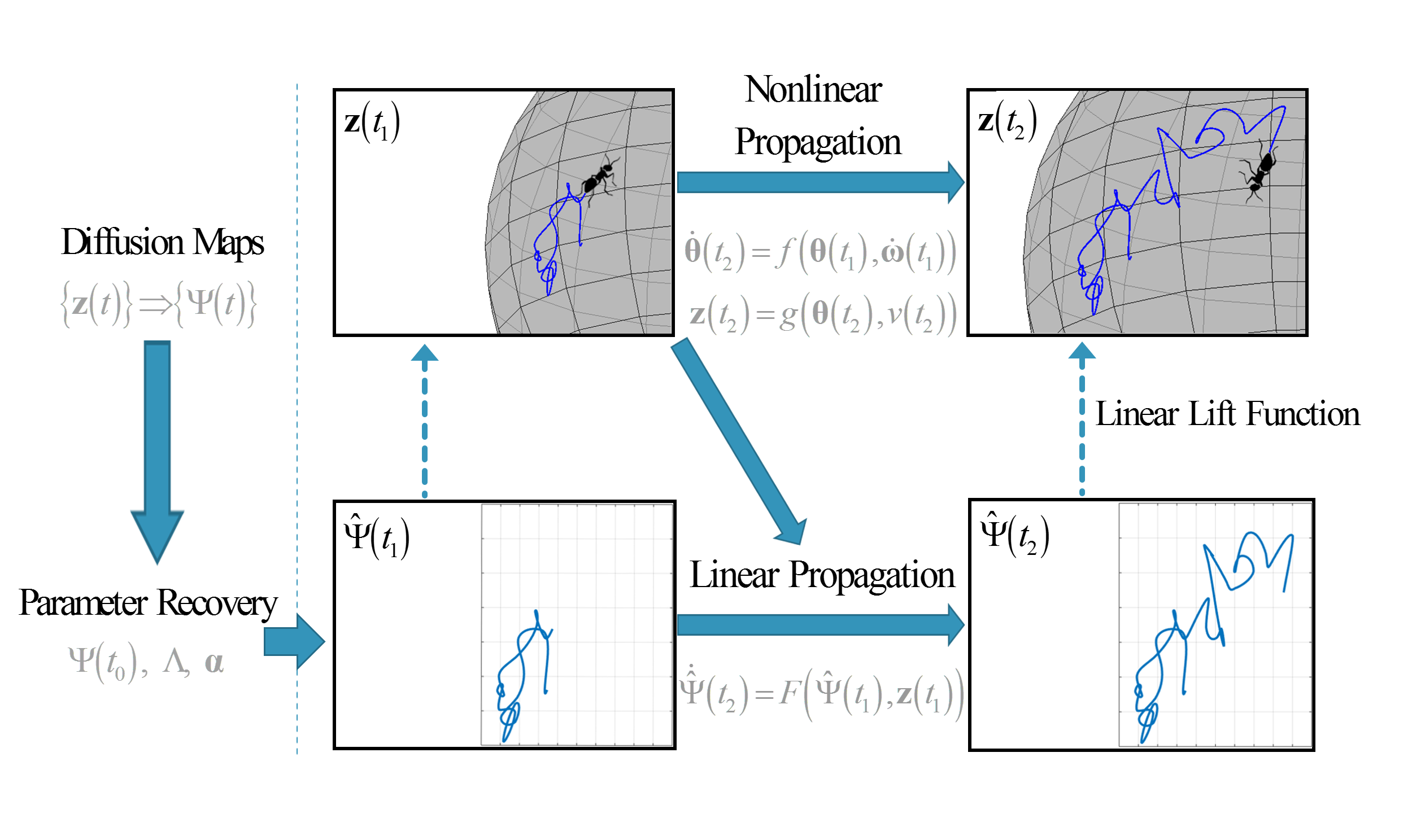}\caption{A schematic view of the observer framework. Similarly to the Koopman operator, the observer provides a parametrization of the underlying state which evolves according to linear dynamics.}
\label{fig:KoopmanFigure} 
\end{figure}

Figure \ref{fig:KoopmanFigure} presents a schematic view of the observer framework. The observer is constructed based on the dynamics parameter $\Lambda$ and lift function $\boldsymbol{\alpha}$, revealed by diffusion maps. Figure \ref{fig:KoopmanFigure} shows that similarly to the Koopman operator, the attained observer propagates according to a linear function \eqref{eq:Observer_actual}, marked by $F\left(\cdot\right)$ in the figure, even for nonlinear systems in which the underlying process and measurements are governed by nonlinear functions \eqref{eq:dyn_sys1},\eqref{eq:dyn_sys2}, marked by $f$ and $g$ in the figure.

\subsection{Implementation in Discrete Setting}

The derivations up to this point are presented for a continuous setting,
however, in most cases only a finite set of discrete measurements
is available. In this section we present a discrete setting in which
the eigenvalues and eigenfunctions of the Fokker-Planck
operator introduced in Section \ref{sec:Recovery-of-System} can be approximated, and we describe
the discrete version of the constructed observer.

Given a set of $N$ measurements, $\left\{ \boldsymbol{z}\left(t_{i}\right)\right\} _{i=0}^{N-1}$,
where $t_{i}=i\cdot\Delta t$, the discrete diffusion maps algorithm
is given in Algorithm \ref{alg:Diffusion maps discrete}.
\begin{algorithm}
\begin{enumerate}
\item Construct the affinity matrix, $\boldsymbol{K}$, 
\[
K\left(i,j\right)=\exp\left(-\frac{d\left(\boldsymbol{z}\left(t_{i}\right),\boldsymbol{z}\left(t_{j}\right)\right)}{\epsilon}\right)
\]
where $d\left(\boldsymbol{z}\left(t_{i}\right),\boldsymbol{z}\left(t_{j}\right)\right)$
is the modified Mahalanobis distance in \eqref{eq:Mahalanibis Eq}.
\item Create the row stochastic matrix: 
\[
W\left(i,j\right)=\frac{1}{D\left(i\right)}K\left(i,j\right)
\]
where $D\left(i\right)=\sum_{j=0}^{N-1}K\left(i,j\right)$. 
\item Compute the eigenvalues, $\mu_{0},...,\mu_{N-1}$, and eigenvectors,
$\psi_{0},...,\psi_{N-1}$, of the matrix $\boldsymbol{W}$. 
\end{enumerate}
\protect\protect\protect\caption{\label{alg:Diffusion maps discrete}Diffusion maps discrete setting}
\end{algorithm}
Based on this algorithm we compute the eigenvalues $\mu_{0},\dots,\mu_{N-1}$
and eigenvectors $\psi_{0},...,\psi_{N-1}$ of the row-stochastic
matrix $\boldsymbol{W}$, and order them such that $1=\mu_{0}\ge\mu_{1}\ge\dots\ge\mu_{N-1}\ge0$.
Note that $\psi_{0}=\begin{bmatrix}1 & 1 & \cdots & 1\end{bmatrix}^{T}$
is the trivial eigenvector associated with $\mu_{0}=1$; the next
few eigenvectors provide a coordinate system for the data, so that
$\psi_{\ell}(i)$, the $i$-th entry of $\psi_{\ell}$, provides the
$\ell$-th coordinate for $\boldsymbol{z}(t_{i})$.

As in Section \ref{sub:Manifold-Learning-and} we view the $m$ eigenvectors
associated with the $m$ largest eigenvalues (except the trivial $\psi_{0}$)
as a parametrization of the state of the system \cite{TalmonPNAS}.
These $m$ eigenvectors form empirical embedding coordinates of the
data by the mapping:
\begin{equation}
\mathbf{z}(t_{i})\mapsto[\psi_{1}(i),\psi_{2}(i),,\psi_{m}(i)]^{T},\ i=0,\ldots,N-1\label{eq:mapping_discrete}
\end{equation}
These eigenvectors are shown to be discrete approximations of the
eigenfunctions of the Fokker-Planck operator $\mathcal{L}$ \eqref{eq:Fokker_Planck} in \cite{Belkin2008,Coifman2006,Singer2006,Nadler2006}.
In addition, the corresponding eigenvalues of the Fokker-Planck operator
can be approximated by $-\lambda_{\ell}=\frac{2}{\beta\epsilon}\log\mu_{\ell}$,
where $\mu_{\ell}$ are the eigenvalues of the row-stochastic matrix
$\boldsymbol{W}$ \cite{Singer2007}. This adaptation is necessary
since the accrued variance in each time step of the stochastic equation
\eqref{eq:StochEq} is $2/\beta$ and the matrix $\boldsymbol{W}$
represents a Markov chain with time steps of $\epsilon$.

Based on these eigenvalues and eigenvectors, the discrete observer
is constructed as described in Algorithm \ref{alg:Discrete-observer-framework}.

\begin{algorithm}
\begin{enumerate}
\item Construct the lift function based on the computed eigenvectors and
given measurements 
\[
\alpha_{j,\ell}=\langle z_{j},\psi_{\ell}\rangle=\sum_{i=0}^{N-1}z_{j}(t_{i})\psi_{\ell}(t_{i})\,,\,\,\ell=1,...,m
\]

\item The discrete observer is given by 
\begin{align}
\widehat{\Psi}\left(t_{i+1}\right) & =\widehat{\Psi}\left(t_{i}\right)+\Lambda\widehat{\Psi}\left(t_{i}\right)+\boldsymbol{\kappa}(\boldsymbol{z}\left(t_{i}\right)-\hat{\boldsymbol{z}}\left(t_{i}\right))\nonumber \\
 & =\left[\mathrm{I}+(1-\gamma)\Lambda\right]\widehat{\Psi}\left(t_{i}\right)+\gamma\Lambda\boldsymbol{\alpha}^{\dagger}\boldsymbol{z}\left(t_{i}\right)\label{eq:discrete_observer}
\end{align}
where $\Lambda$ is a diagonal matrix with $\Lambda_{\ell\ell}=-\lambda_{\ell}$,
$\ell=1,...,m$, $\mathrm{I}$ is the identity matrix and $\boldsymbol{\kappa}$
is chosen as in \eqref{eq:Observer_actual}. 
\end{enumerate}
\caption{\label{alg:Discrete-observer-framework}Discrete observer framework}
\end{algorithm}

For simplicity, the discrete observer \eqref{eq:discrete_observer}
is presented with $\Delta t=t_{i+1}-t_{i}$. However, we note that
in general, the time step can be altered from the time step of the
given measurements, $\Delta t=t_{i+1}-t_{i}$, by multiplying $\Lambda\widehat{\Psi}\left(t_{i}\right)+\boldsymbol{\kappa}(\boldsymbol{z}\left(t_{i}\right)-\hat{\boldsymbol{z}}\left(t_{i}\right))$
with the desired $\Delta t^{*}<\Delta t$ and fixing the measurement
at time $t_{i}$ for all $t_{i}+n\Delta t^{*}<t_{i+1}$, $n\in\mathbf{N}$.

\subsection{Diffusion Filtering}

In the discrete setting, the interpretation in \eqref{eq:Observer_actual}
can be extended further by using the approximation \eqref{eq:approx_g},
which yields: 
\begin{equation}
\widehat{\Psi}\left(t_{i+1}\right)-\widehat{\Psi}\left(t_{i}\right)=(1-\gamma)\Lambda\widehat{\Psi}\left(t_{i}\right)+\gamma\Lambda\Psi\left(t_{i}\right)\label{eq:observer_Psi}
\end{equation}

We now see that the evolution in time of the observer's estimated state is a weighted
sum of $\Psi$ and $\hat{\Psi}$ controlled by $\gamma$. Since the
eigenvectors can be viewed as solutions of the Fokker-Planck equation,
they describe the propagation in time of the density evolution due
to diffusion. Accordingly, we can interpret the two terms in the recursive
equation. Both terms $\Lambda\Psi$ and $\Lambda\hat{\Psi}$ can be
viewed as an approximation of the propagation of the probability density
of $\boldsymbol{\theta}\left(t\right)$ one step forward as presented in \eqref{eq:eigv_dyn}.
On the one hand, $\Lambda\Psi$ is one step forward from $\Psi$ (which
was constructed merely from the data and is independent of the time
sequence). On the other hand, $\Lambda\hat{\Psi}$ is one step forward
from $\hat{\Psi}$ which is a ``trajectory of evolving densities''
from the initial point, and hence, it is time dependent.

To make the explanation above more clear, we write the differential
equation \eqref{eq:observer_Psi} explicitly in the discrete form
by recursively telescoping $\hat{\Psi}\left(t_{i}\right)$, which
yields
\begin{align}
\widehat{\Psi}\left(t_{n+1}\right) & =\left[\mathrm{I}+(1-\gamma)\Lambda\right]^{n+1}\Psi\left(t_{0}\right)+\nonumber \\
 & +\gamma\Lambda\sum_{i=0}^{n}\left[\mathrm{I}+(1-\gamma)\Lambda\right]^{(n-i)}\Psi\left(t_{i}\right)\label{eq:observer_filter}
\end{align}
where $\widehat{\Psi}\left(t_{0}\right)=\Psi(t_{0})$.

The explicit form in \eqref{eq:observer_filter} highlights the two
terms comprising an effective filtering applied by the observer.
The first term is the propagation of the density $n+1$ steps forward
from the initial point. This term encapsulates merely the diffusion
propagation of the densities as captured by diffusion maps (with $\gamma$
weighting), such that it is enhanced by small values of $\gamma$.
However, it does not explicitly take into account the samples along
the given trajectory. For example, for $\gamma=0$ we have 
\begin{equation}
\widehat{\Psi}(t_{n+1})=(\mathrm{I}+\Lambda)^{n+1}\Psi(t_{0})
\end{equation}
The second term is the propagation of a single step forward from each
point along the trajectory. This term can be viewed as the correction
of the general propagation of the densities, stemming from the particular
realization at hand. The term is enhanced for large values of $\gamma$,
for example, for $\gamma=1$ we have 
\begin{equation}
\widehat{\Psi}(t_{n+1})=\Psi(t_{0})+\Lambda\sum_{i=0}^{n}\Psi(t_{n})
\end{equation}

We note that \eqref{eq:observer_filter} takes the form of a filter:
$\gamma\Lambda\Psi(t_{n})$ convolved with the filter $[\mathrm{I}-(1-\gamma)\Lambda)]^{n}$,
which weighs past samples exponentially.

Moreover, if for simplicity we initiate the observer with zero and
place $\Psi\left(t_{i}\right)=\boldsymbol{\alpha}^{\dagger}\boldsymbol{z}\left(t_{i}\right)$,
the observer can be written as a data driven, moving average filter
on the transformed measurements, $\boldsymbol{\alpha}^{\dagger}\boldsymbol{z}$,
in which the window size is determined by the dynamics coefficient,
$\Lambda$, and the parameter $\gamma$: 
\begin{equation}
\widehat{\Psi}\left(t_{n+1}\right)=\gamma\Lambda\sum_{i=1}^{n}\left[\mathrm{I}+(1-\gamma)\Lambda\right]^{(n-i)}\boldsymbol{\alpha}^{\dagger}\boldsymbol{z}\left(t_{i}\right)\label{eq:DiffusionFilter_Z}
\end{equation}

\subsection{Observer Based Out-of-Sample Extension\label{sub:Observer-Based-Out-of-Sample}}

One of the main shortcomings of addressing dynamical systems using
manifold learning techniques concerns the handling of streaming data.
When a new measurement $z\left(t_{i}\right)$ is acquired, we wish
to extend the learned coordinate system. This is commonly performed
by the Nystr\"{o}m extension \cite{bengio2004Nyst}, which is an extension
scheme for eigenvectors $\psi_{\ell}$: 
\begin{equation}
\psi_{\ell}\left(t_{i}\right)=\frac{1}{\lambda_{\ell}}\sum_{j=0}^{N-1}W\left(i,j\right)\psi_{\ell}\left(t_{j}\right)\label{eq:Nystrom}
\end{equation}
where $\boldsymbol{W}$ is constructed as described in Algorithm \ref{alg:Diffusion maps discrete},
based on the affinity kernel between the new measurement $\boldsymbol{z}\left(t_{i}\right)$
and the measurements with known eigenvectors. 

In our case, the new coordinate system constructed by diffusion maps
consists of eigenvectors, and therefore, the Nystr\"{o}m extension can be applied.
This extension allows extrapolation of the learned parametrization
to new, unseen measurements. However, it is accurate only for measurements
which are closely related to known data. In this section we present
an out-of-sample extension scheme which naturally arises from the
observer's structure.

As presented in Section \ref{sub:Fixed-Contractive-Observer}, the
observer is constructed based on the dynamics matrix $\Lambda$ and
the lift function $\boldsymbol{\alpha}$, both revealed by the diffusion
maps framework. We assume that the dynamics and reconstruction function
do not change significantly in time and therefore describe the system
for new measurements as well. We present the proposed out-of-sample
extension framework in Algorithm \ref{alg:Out-of-sample-Extension}. 

\begin{algorithm}
\begin{enumerate}
\item Given an initial set of $N-1$ measurements, apply diffusion maps
to reveal system dynamics $\Lambda_{N-1}$ and reconstruction matrix
$\boldsymbol{\alpha}_{N-1}$. 
\item Construct the observer for the initial set of measurements. 
\item Given a new measurement $\boldsymbol{z}\left(t_{N}\right)$ apply 
\begin{eqnarray}
\widehat{\Psi}\left(t_{N+1}\right) & = & \left[\mathrm{I}+(1-\gamma)\Lambda_{N-1}\right]\widehat{\Psi}\left(t_{N}\right)+\nonumber \\
 &  & +\gamma\Lambda_{N-1}\boldsymbol{\alpha}_{N}^{\dagger}\boldsymbol{z}\left(t_{N}\right)
\end{eqnarray}

\end{enumerate}
\protect\caption{\label{alg:Out-of-sample-Extension}Out-of-sample Extension}
\end{algorithm}

\section{Experimental Results\label{sec:Experimental-Results}}

\subsection{Demonstrating the Estimation of the Underlying Process\label{sub:Estimation-of-the}}

We base our toy example on the setting presented in \cite{TalmonPNAS}
which describes a radiating object moving on a 3D sphere. The process
is described by its elevation and azimuth angles, as a 2D Langevin
equation with a parabolic potential: 
\begin{eqnarray*}
\dot{\theta}_{1} & = & \left(\frac{\pi}{2}\cdot c-c\cdot\theta_{1}\right)+b\dot{\omega}_{1}\\
\dot{\theta}_{2} & = & \left(\frac{\pi}{10}\cdot c-c\cdot\theta_{2}\right)+b\dot{\omega}_{2}
\end{eqnarray*}
where $b$ is the diffusion coefficient (set to 0.005) and $c$
is the drift rate parameter which we vary between 0.08 and 0.024.

The resulting 3D process is given by: 
\begin{eqnarray*}
x_{1}\left(t\right) & = & \cos\left(\theta_{2}\right)\sin\left(\theta_{1}\right)\\
x_{2}\left(t\right) & = & \sin\left(\theta_{2}\right)\sin\left(\theta_{1}\right)\\
x_{3}\left(t\right) & = & \cos\left(\theta_{1}\right)
\end{eqnarray*}

\begin{figure}[t]
\begin{centering}
\includegraphics[width=0.7\columnwidth]{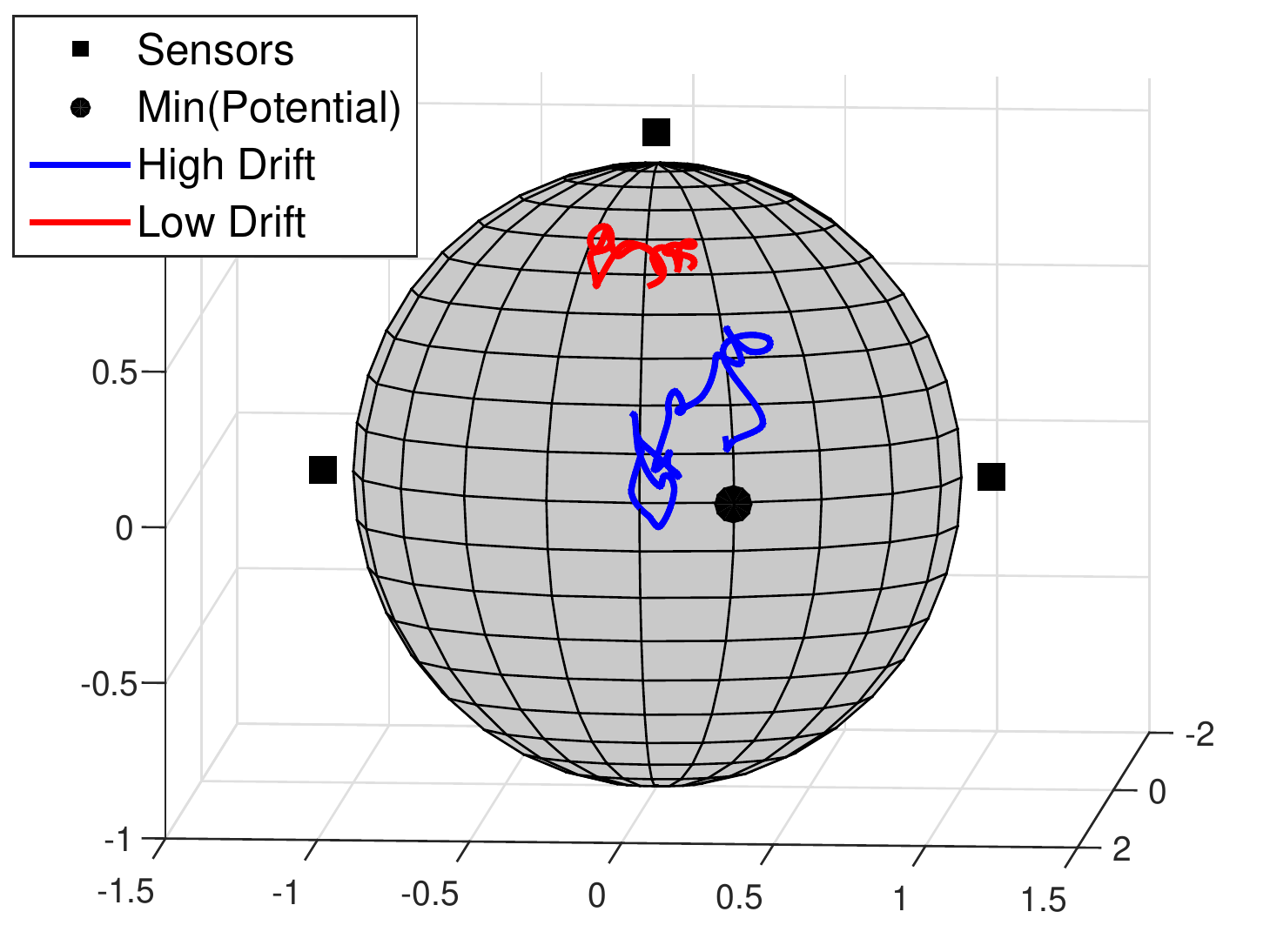} 
\par\end{centering}

\begin{centering}
\includegraphics[width=0.7\columnwidth]{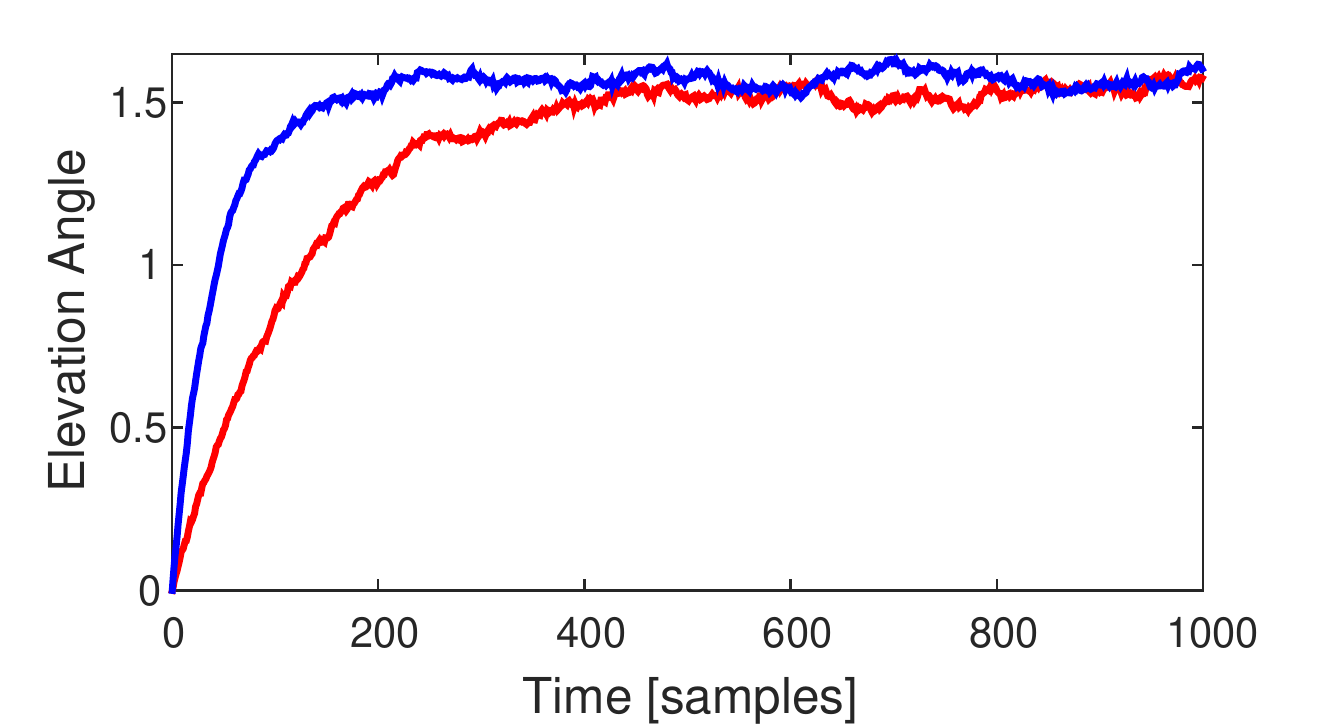} 
\par\end{centering}

\protect\protect\protect\protect\caption{Toy example setup with different drift rates $c$. (top) Two 300 point
segments of the 3D movement on the sphere with different
drift rates. (bottom) Two realizations of the elevation angle over
$1000$ time samples with different drift rates. The red segments
corresponds to a process with $c=0.008$ and the blue segments corresponds to a process with $c=0.024$.}
\label{fig:PNAS_DifferentDrift_Sphere_Example} 
\end{figure}

We mark the 3D location of the object at time $t$ by $\boldsymbol{x}\left(t\right)=\left[x_1\left(t\right),x_2\left(t\right),x_3\left(t\right)\right]$.
The position of the object is measured by 3 sensors located at $\boldsymbol{s}_{1},\boldsymbol{s}_{2},\boldsymbol{s}_{3}$
as presented in Figure \ref{fig:PNAS_DifferentDrift_Sphere_Example}.
Each sensor $\boldsymbol{s}_j$ is modeled as a ``Geiger Counter'' and fires spikes
according to a Poisson distribution in a rate, $r_{j}$, which depends
on the proximity of the object to the sensor: $r_{j}\left(t\right)=\exp\left(-\left\Vert \boldsymbol{s}_{j}-\boldsymbol{x}\left(t\right)\right\Vert \right)\,\,,\,j=1,2,3$.

The output of each sensor is described by: 
\begin{eqnarray*}
z_{j}\left(t\right) & = & y_{j}\left(t\right)+v_{j}\left(t\right)\\
y_{j}\left(t\right) & \sim & Pois\left(r_{j}\left(t\right)\right)\,\,\,j=1,2,3
\end{eqnarray*}
where $v_{j}\left(t\right)$ is a spike train drawn from a Poisson
distribution with a fixed rate parameter. Here, $y_{j}\left(t\right)$
represents the measurement modality of the intrinsic process and $v_{j}\left(t\right)$
represents an additive, independent measurement noise. The available
measurements $z_{j}\left(t\right)$ are the sum of $y_{j}\left(t\right)$
and $v_{j}\left(t\right)$.

Our goal in this example is to recover the underlying 2D dynamical
process of the angles based solely on the measurements in this non-linear,
non-Gaussian setting.

In this section the diffusion maps framework is implemented similarly
to the one described in \cite{TalmonPNAS}. In this framework, histograms
are constructed as feature vectors and are shown to estimate the probability
density function of the clean observation process $y_{i}\left(t\right)$.
The histograms are calculated based on non-overlapping time frames,
each containing $60$ time samples, $z_{j}\left(t_{i}:t_{i+59}\right)$.
After histogram construction, the Mahalanobis distance, presented
in \eqref{eq:Mahalanibis Eq}, is calculated based on these histograms.
Mahalanobis distance is then used to construct the pairwise affinity
kernel as described in \eqref{eq:PairwiseAffinity}. Diffusion maps
coordinates and the observer's estimated state are constructed based on the description
in Section \ref{sub:Manifold-Learning-and} and Section \ref{sec:The-Observer}.
To summarize, the analysis is performed as follows: 
\begin{enumerate}
\item Given $N$ measurements, construct histograms based on time frames
of 60 samples, resulting in $\frac{N}{60}$ histograms. 
\item Calculate the Mahalanobis distance based on these $\frac{N}{60}$
histograms and build a pairwise affinity matrix of size $\frac{N}{60}\times\frac{N}{60}$. 
\item Continue as described in Algorithm \ref{alg:Diffusion maps discrete}
and Algorithm \ref{alg:Discrete-observer-framework}. 
\end{enumerate}
Note that the first step of Algorithm \ref{alg:Diffusion maps discrete}
requires knowledge of $\epsilon$. Common practice is to set the scale
$\epsilon$ to be of the order of the median of the pairwise distances $d\left(\boldsymbol{z}\left(t_{i}\right),\boldsymbol{z}\left(t_{j}\right)\right)$,
$\forall i,j$. Here, we set it to be the median itself, since empirically it was shown to attain good performance.

We generated $240,000$ time samples of this process in time steps
of $\Delta t=0.1$. As described above, this correspond to $4000$
histograms of $z_{j}\left(t\right)$ and therefore to $4000$ recovered
coordinates (embedding coordinates). The measurement modality and diffusion maps
construction is performed as described at the beginning of this section
and the observer is created based on the eigenvalues and eigenvectors
as described in Section \ref{sec:The-Observer}. We applied the proposed
observer (with $\gamma=0.85$) and compared its performance to the
diffusion maps embedding in recovering the underlying 2D dynamical
process, $\theta_{i}\left(t\right)$.

In order to evaluate our results more accurately, we initially perform
linear regression on the first 4 diffusion maps coordinates. This
is carried out since one of the shortcomings of the diffusion maps
coordinates is that, while they can provide a good representation
of the underlying state, they are not necessarily separable \cite{dsilva2015parsimonious}. Therefore, the linear regression is performed merely for
the purpose of comparing between representations of the underlying state by the
observer and by the diffusion maps coordinates. We note that without
this linear regression the representation errors are larger, however,
the observer still improves the representation of the underlying state,
compared with the diffusion maps coordinates.

\begin{figure}[t]
\centering{}\includegraphics[width=1\columnwidth]{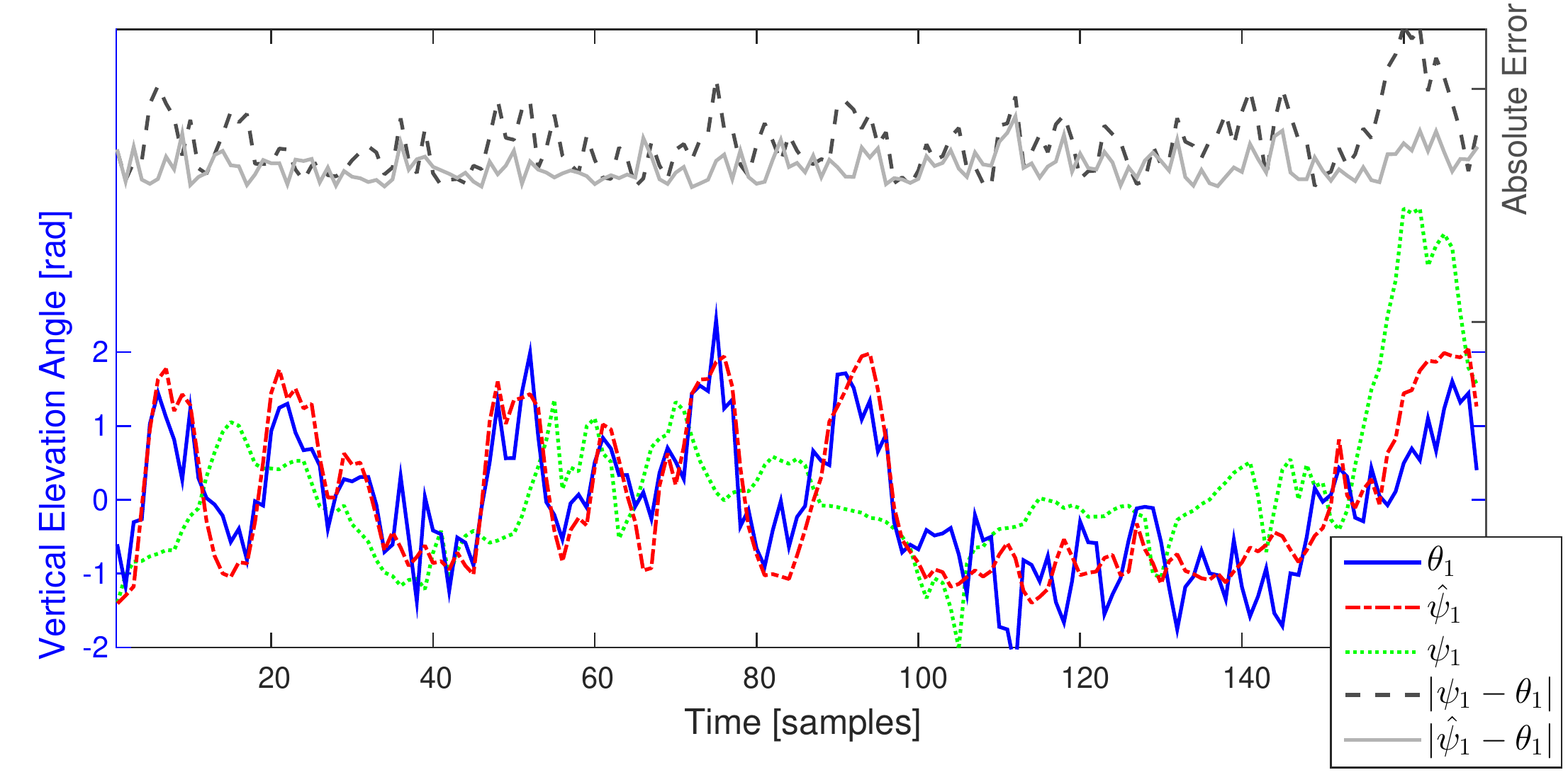}\protect\caption{Vertical elevation angle recovery. Comparison between the true vertical
angle (blue line), created with $c=0.024$, the first coordinate
of the constructed observer (dashed red line) and the diffusion maps
embedding (dotted green line). The absolute errors of the observer
coordinate (light gray line) and the diffusion maps coordinate (dashed
gray line) are presented at the top of the plot.}
\label{fig:PNAS_Time_Series} 
\end{figure}

In Figure \ref{fig:PNAS_Time_Series}, a short interval of the normalized
and centered vertical elevation angle process is displayed (blue plot),
along with the first diffusion map coordinate (green plot) and the
first coordinate of the observer's estimated state (red plot). Each point in the plots of the observer coordinate and diffusion maps coordinate represents one histogram of
$60$ time samples and therefore, the true angle, $\theta_{1}\left(t\right)$,
is down-sampled. At the top of the figure, the absolute errors between
the true elevation angle and both coordinates are displayed in gray. Note that we refer to the recovered state based on the observer equation, as the observer coordinates, in order to avoid confusion with the recovered state based on diffusion maps.

It is noticeable in Figure \ref{fig:PNAS_Time_Series} that both coordinates
follow the general trend of the true angle, however the observer coordinate represents
the true angle more accurately as can be seen in the coordinate
plots and in the absolute error plot. 

\begin{figure}[t]
\centering{}
\subfloat[]{\protect\includegraphics[width=0.21\textwidth]{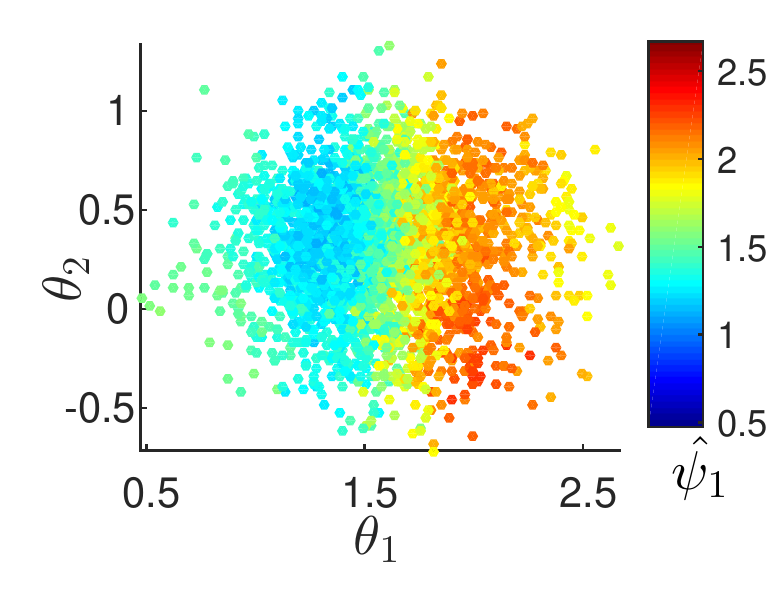}

}\subfloat[]{\protect\includegraphics[width=0.21\textwidth]{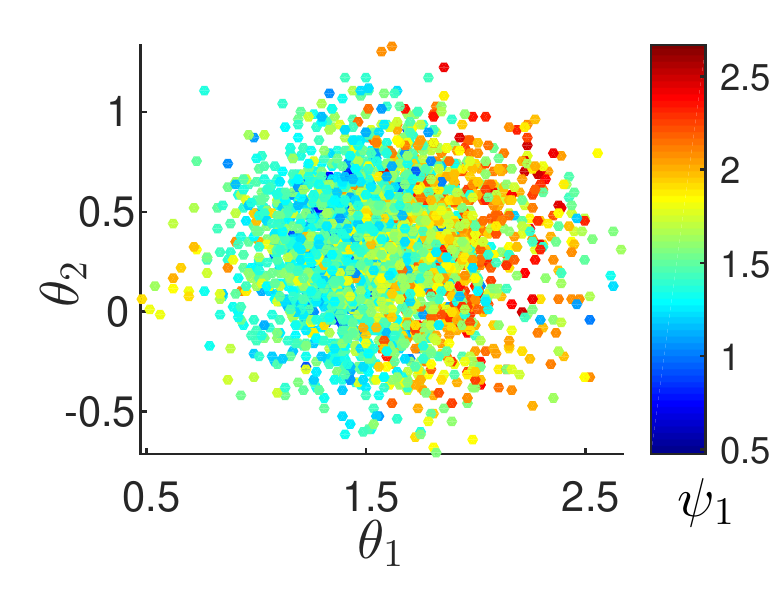} }

\vspace{-0.4cm}
\subfloat[]{\protect\includegraphics[width=0.21\textwidth]{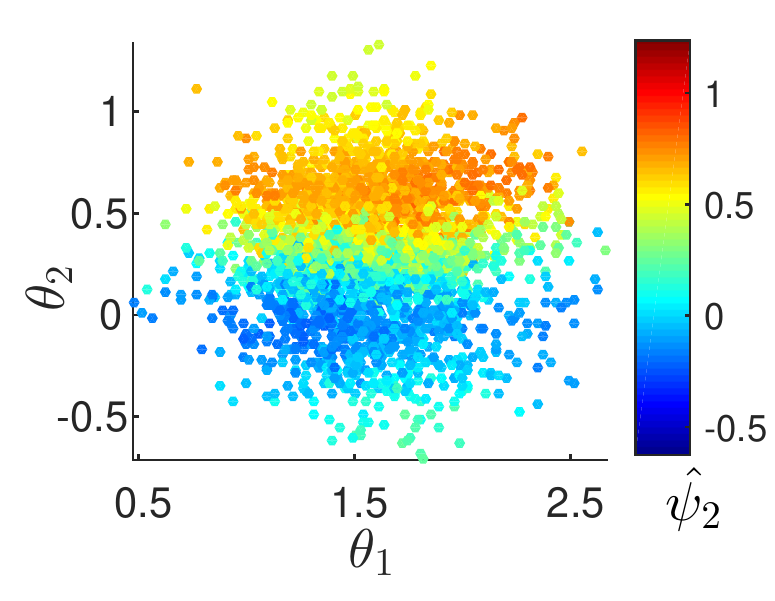}

}\subfloat[]{\protect\includegraphics[width=0.21\textwidth]{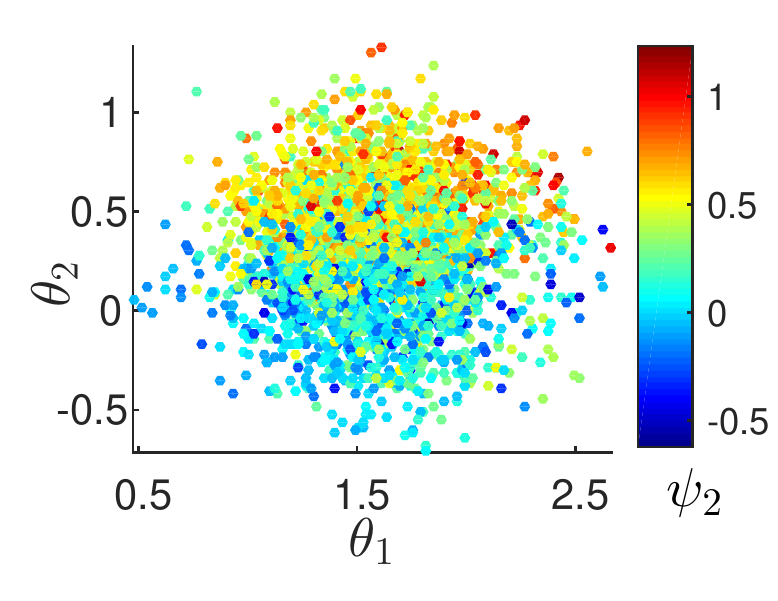} }
\vspace{-0.2cm}

\caption{Scatter plots of the vertical and horizontal angles. The plots are
colored by values of the first and second observer coordinates (plots
a,c) and the values of the first and second diffusion maps coordinates
(plots b,d). The angles were simulated with a drift rate parameter
of $c=0.024$.}
\label{fig:PNAS_Scatter_Correlation} 
\end{figure}

This improvement in correlation is emphasized in Figure \ref{fig:PNAS_Scatter_Correlation}.
The figure contains four identical scatter plots of the two underlying
angles i.e., $\theta_{2}\left(t\right)$ as a function of $\theta_{1}\left(t\right)$,
which differ only in their color schemes. Each of the four plots (denoted
by (a),(b),(c),(d)) is colored based on either the first and second observer
coordinates (plots (a) and (c) respectively) or the first and second diffusion
maps coordinates (plots (b) and (d) respectively). The color schemes of these
scatter plots depict that both underlying angles are represented better
by the first and second observer coordinates, since the color gradient
is smoother in plots (a) and (c), than the color gradient in plots (b) and (d). Therefore, as expected in the presented toy example which suffers from noise,
both additive and model based (due to the stochastic nature of the
sensors), the observer describes the underlying dynamical process
more accurately.

We note that although the observer provides better representations
of the underlying state than the diffusion maps coordinates, it suffers
from inaccuracies at the boundaries of the data. This is visible in Figure
\ref{fig:PNAS_Scatter_Correlation}, for example in plot (a) when $\theta_{1}<1$.
This is due to the inaccuracy of the linear lift function at the boundaries and will be addressed in future work.

As described in \eqref{eq:DiffusionFilter_Z}, the observer is a data-driven
filter applied to $\boldsymbol{\alpha}^{\dagger}\boldsymbol{z}$ with
varying filter length which is determined by $\Lambda$ and $\gamma$.
Therefore, we compared the result of the observer (with a constant parameter
$\gamma$) to several moving average filters with varying window sizes applied to $\boldsymbol{\alpha^{\dagger}}\boldsymbol{z}$,
in different parabolic potentials (with different parameter $c$ values). Our motivation for varying the drift rate arises from Figure \ref{fig:PNAS_DifferentDrift_Sphere_Example}
which displays the effect of different parabolic potentials on the
resulting path. In this figure, the top plot displays two exemplary
paths on the 3D sphere, one with a high drift rate $c=0.024$ (blue
plot) and one with a low drift rate $c=0.008$ (red plot). The sensor
locations are marked with black rectangles and the location of the
minimum of the parabolic potential is denoted by a black circle. The
bottom plot displays the first coordinate of the 2D underlying state
(elevation angle) for both drift rates. These plots depict the differences
in convergence rate and in the step size, mainly before convergence.
The high drift rate process converges faster, causing higher step
sizes until convergence and its perturbations from the minimum potential
are smaller.

\begin{figure}[t]
\centering{}\includegraphics[width=0.5\columnwidth]{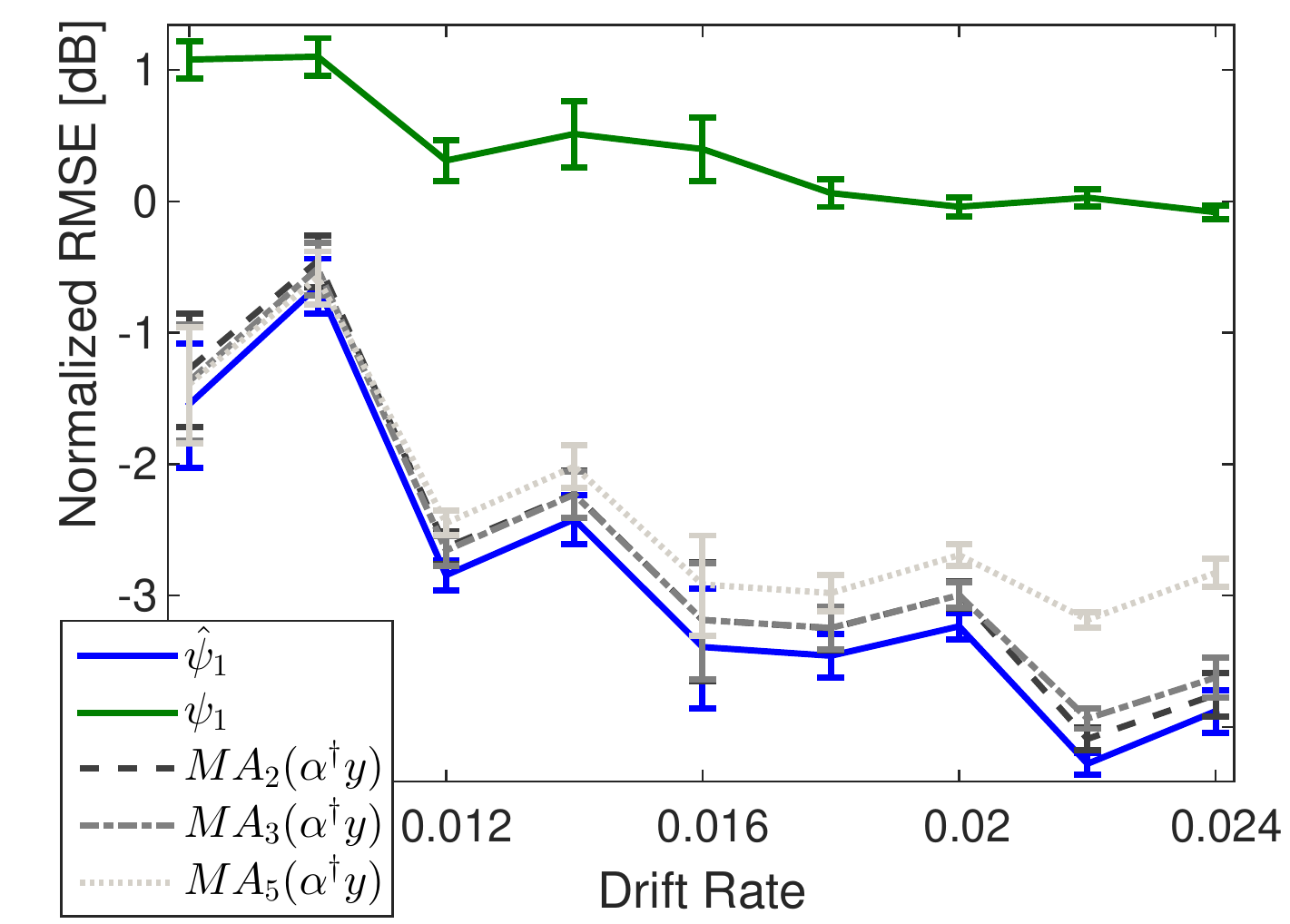}\includegraphics[width=0.5\columnwidth]{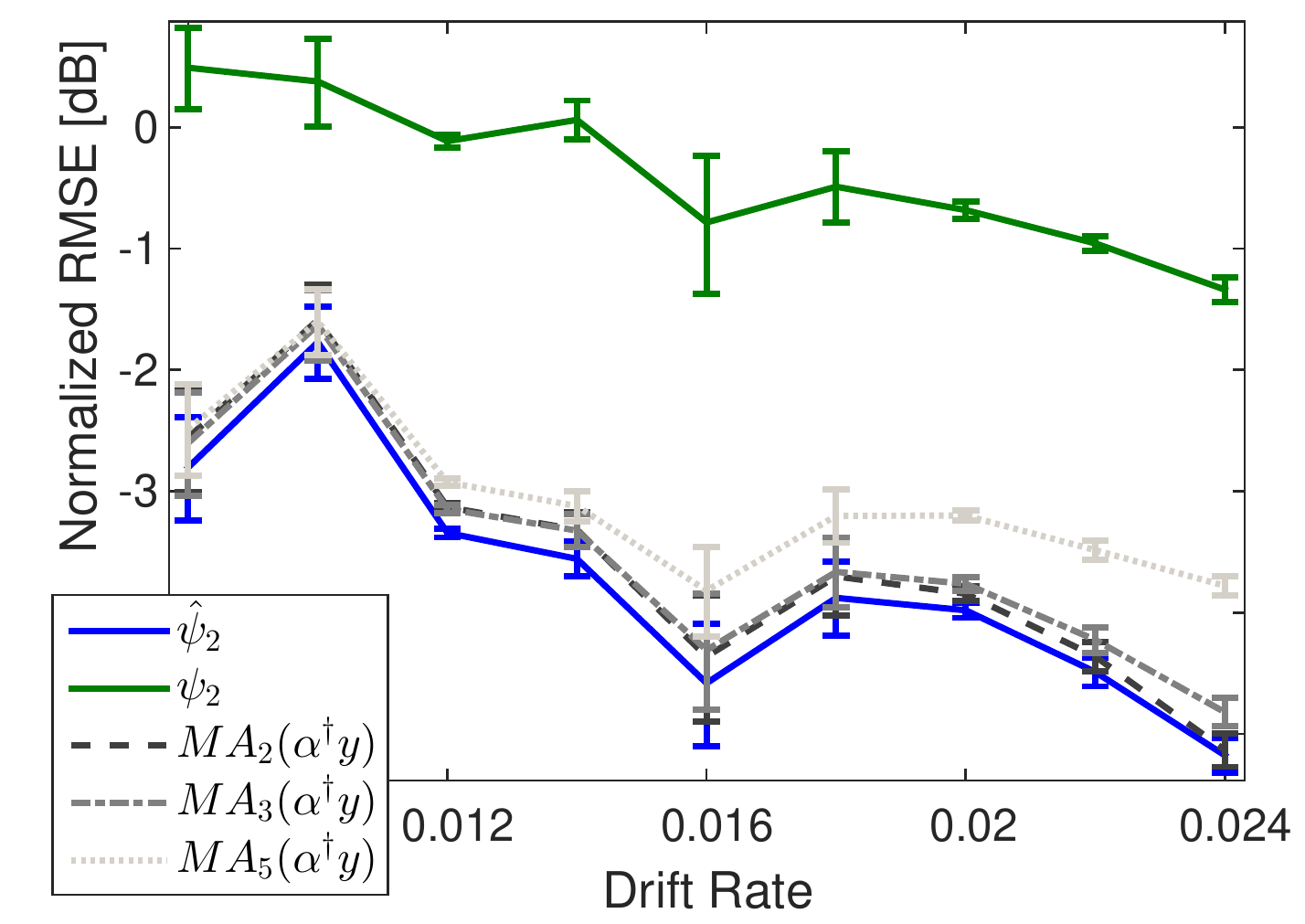}\protect\protect\protect\caption{Normalized RMSE {[}dB{]} between the true angles and their estimations
in different drift rates. The angles are represented by the diffusion
maps embedding (green), observer coordinates (blue) and moving average
filters on \textbf{$\boldsymbol{\alpha}^{\dagger}\boldsymbol{z}$}
with window sizes of 2 (dashed gray), 3 (dot-dashed gray) and 5 (dotted
gray) samples.}
\label{fig:PNAS_plot-nRMSE} 
\end{figure}

\begin{figure}[t]
\centering{}\includegraphics[width=0.5\columnwidth]{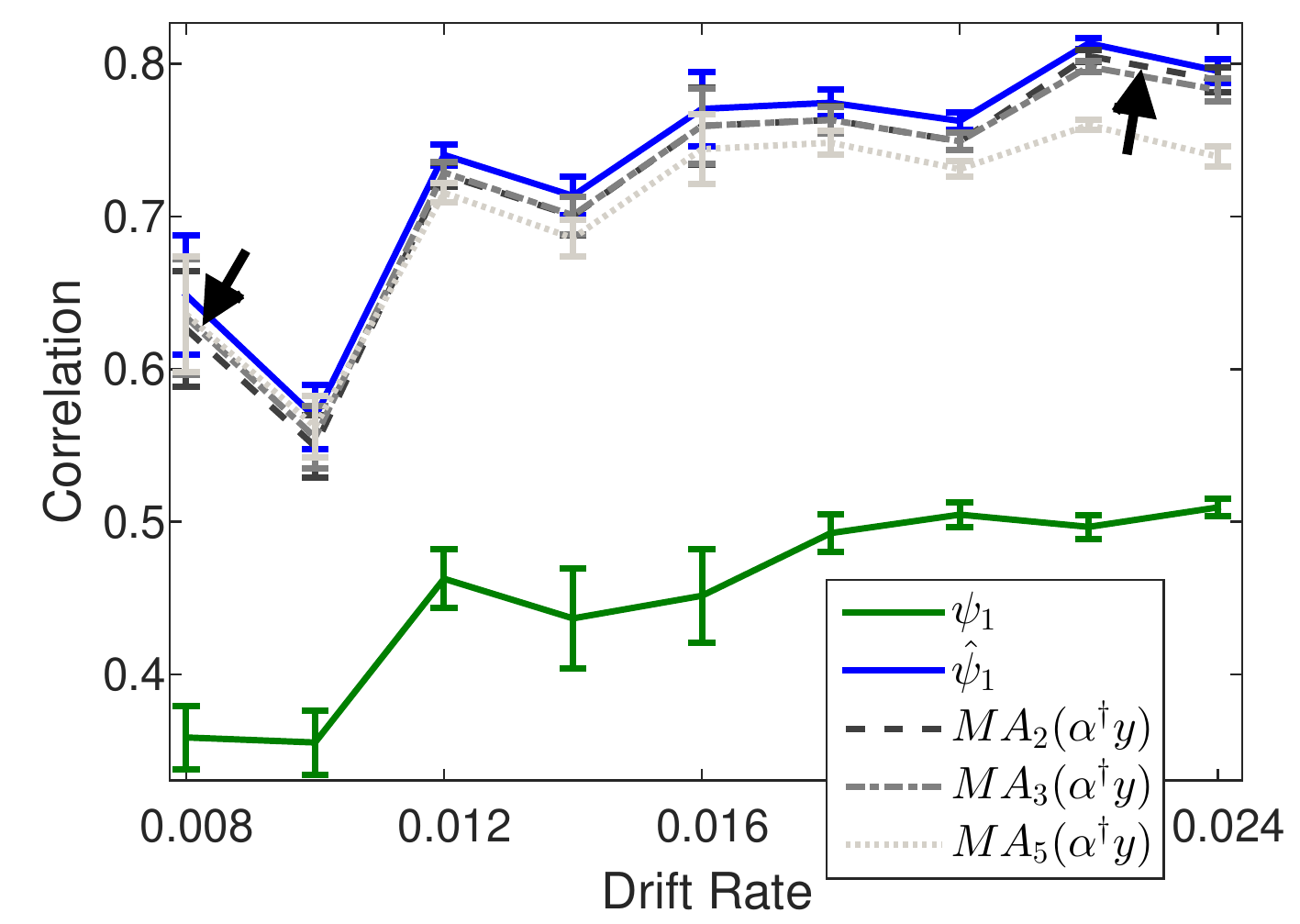}\includegraphics[width=0.5\columnwidth]{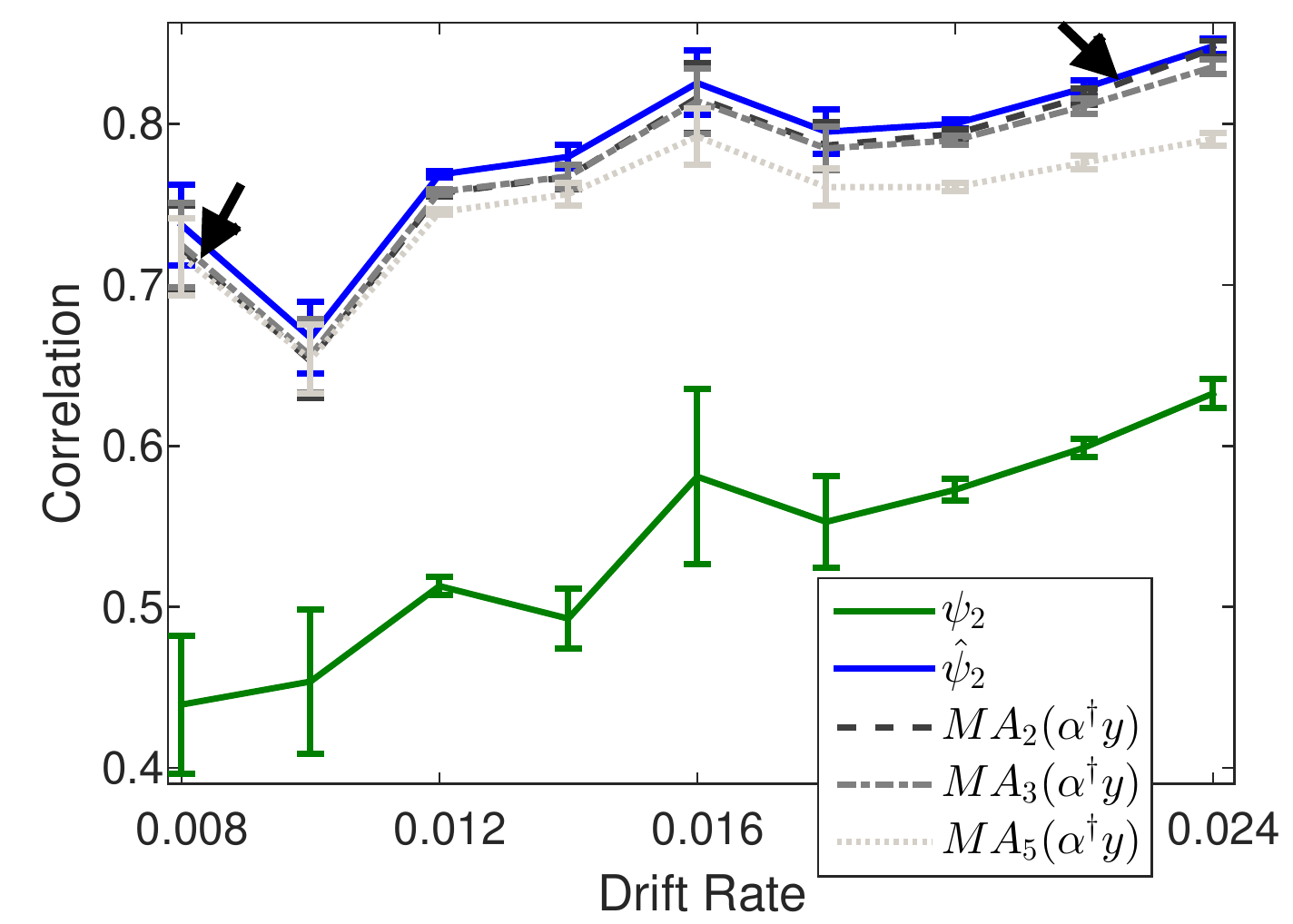}\protect\protect\protect\caption{Correlation between the true angles and their estimations in different
drift rates. The angles are represented by the diffusion maps embedding
(green), observer coordinates (blue) and moving average filters on
\textbf{$\boldsymbol{\alpha}^{\dagger}\boldsymbol{z}$} with window
sizes of 2 (dashed gray), 3 (dot-dashed gray) and 5 (dotted gray).}
\label{fig:PNAS_plot} 
\end{figure}

Figure \ref{fig:PNAS_plot-nRMSE} shows the normalized RMSE, between
the true angles and their estimates, where the plot on the left presents
the the first coordinate (elevation angle) and the plot on the right
presents the second coordinate (azimuth angle). These plots contain
average values of the normalized RMSE, over 50 iterations, with error
bars of one standard deviation, for varying drift rates between $c=0.008$
and $c=0.024$. We compared estimates based on the observer (blue
line), diffusion maps embedding (green line) and moving average filters
on $\boldsymbol{\alpha}^{\dagger}\boldsymbol{z}$ with 3 windows of
lengths $2$,$3$, and $5$ (dashed gray lines). Similarly, Figure
\ref{fig:PNAS_plot} presents the average correlation between the
true angles and their estimates, over 50 realizations, with error
bars of one standard deviation.

It is clear from these plots that the observer's estimated state and moving average
filters outperform the diffusion maps embedding, as their correlation
to the true angles is significantly higher. In addition, it is worth
noting that different averaging filters perform best in different
settings. This is due to the change in path characteristics as the
drift rate parameter $c$ varies, e.g. slow drift rates (wide parabolic
potential) correspond to slower convergence rates and smaller step
sizes, and therefore, a wider averaging filter performs best. Examples
of this effect are marked by black arrows in Figure \ref{fig:PNAS_plot},
where at a low drift rate ($c=0.008$) the moving average filter with
window size of 5 samples performs best (out of the 3 moving average
filters) and at a high drift rate, ($c=0.024$), the shortest moving
average filter performs best.
In all instances, the proposed observer closely follows the best filter
in each setting, demonstrating its superiority as a data driven filter
in which the window size is determined based on the given measurements without prior information, particularly without information on the drift.

\begin{figure}[t]

\centering

\subfloat[]{\protect\includegraphics[width=0.16\textwidth]{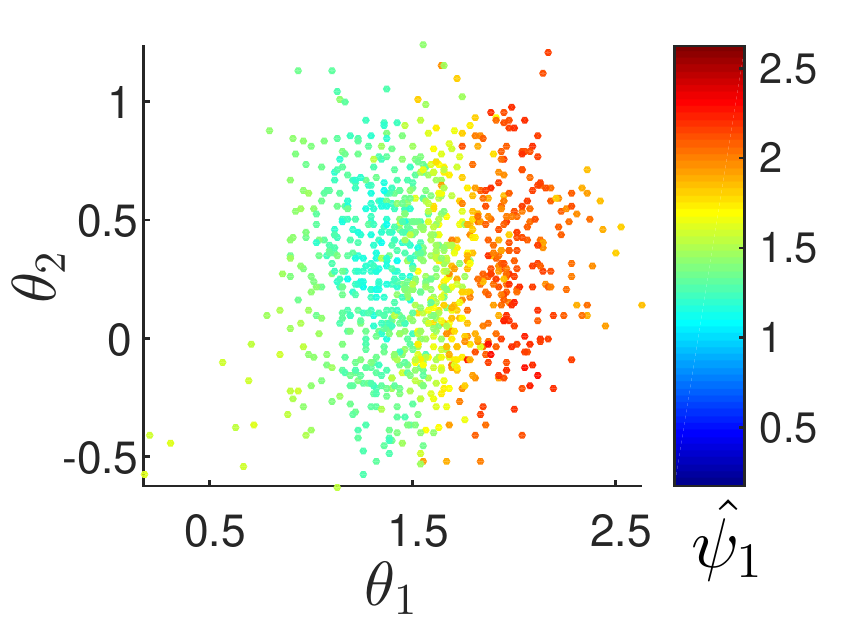}

}\subfloat[]{\protect\includegraphics[width=0.16\textwidth]{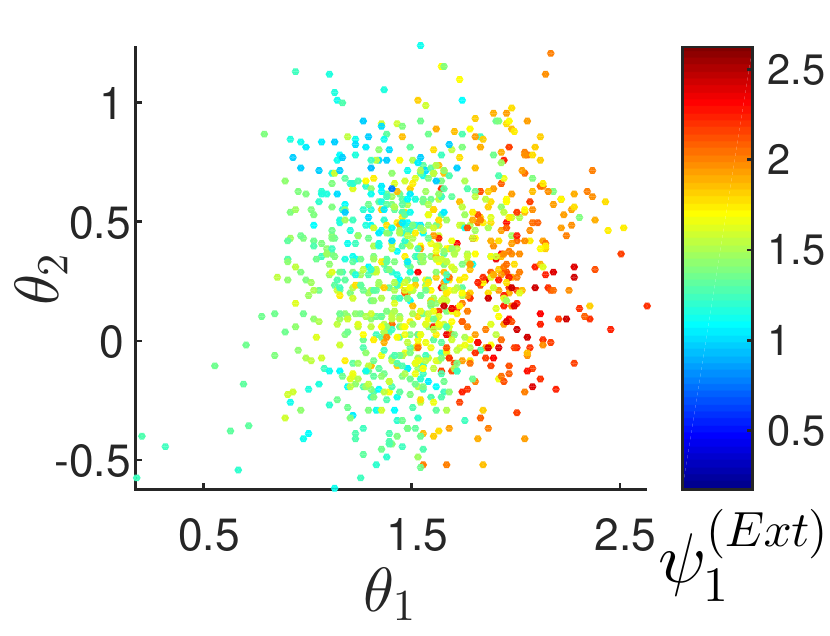}

}\subfloat[]{\protect\includegraphics[width=0.16\textwidth]{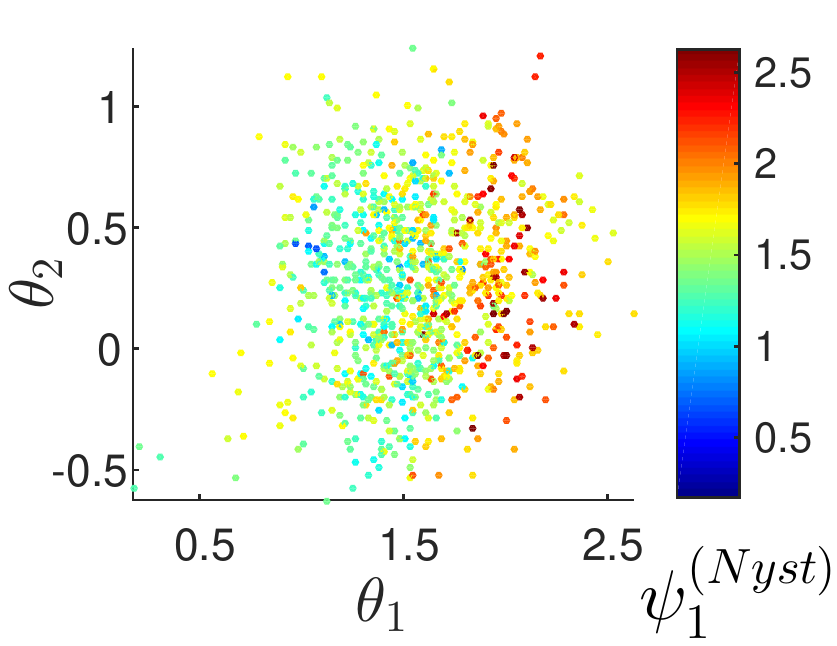} }

\vspace{-0.4cm}
\subfloat[]{\protect\includegraphics[width=0.16\textwidth]{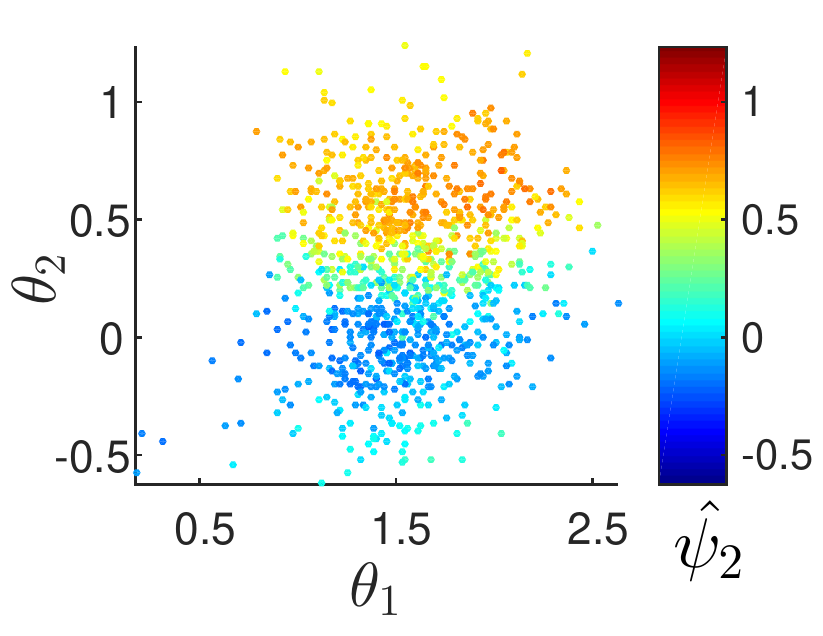}

}\subfloat[]{\protect\includegraphics[width=0.16\textwidth]{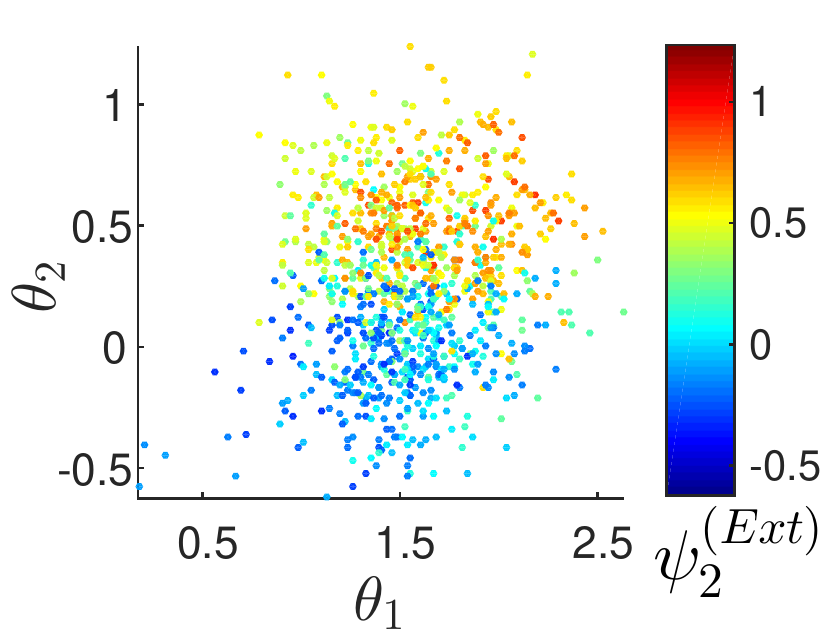}

}\subfloat[]{\protect\includegraphics[width=0.16\textwidth]{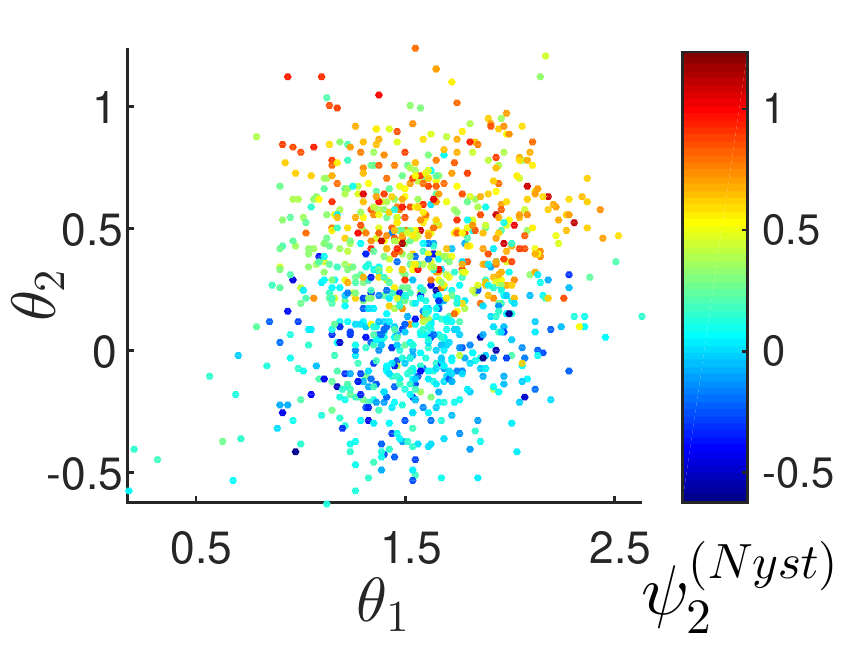} }
\vspace{-0.2cm}
\protect\caption{Extension scheme: Scatter plots of the vertical and horizontal angles,
colored by the extended coordinates. The plots are colored by values
of the first and second observer extension coordinates (plots a,d),
the values of the first and second diffusion maps extension coordinates
in \cite{TalmonPNAS} (plots b,e) and the first and second coordinates
of the Nystr\"{o}m extension (plots c,f). The angles were simulated with
a drift rate parameter of $c=0.024$.}
\label{fig:PNAS_Extension} 
\end{figure}

Finally, we examine the extension scheme presented in Subsection \ref{sub:Observer-Based-Out-of-Sample}.
We simulated a trajectory of length $240,000$, corresponding to $4000$
histograms. We constructed the diffusion maps coordinates and recovered
the required observer parameters, i.e. dynamics and reconstruction
function, based on $3000$ of these histograms as described above.
We then applied the observer extension scheme, as described in Algorithm
\ref{alg:Out-of-sample-Extension}, to the remaining $1000$ histograms,
in order to estimate the 2D underlying process at these time points.
We compared our results to the Nystr\"{o}m extension \eqref{eq:Nystrom}
as a baseline and to the extension scheme proposed in \cite{TalmonPNAS}.
Figure \ref{fig:PNAS_Extension} presents six identical scatter plots
containing these remaining $1000$ time frames (histograms) of the
two angles i.e., $\theta_{2}\left(t\right)$ as a function of $\theta_{1}\left(t\right)$,
which differ only in their color schemes, similarly to Figure \ref{fig:PNAS_Scatter_Correlation}.
In this figure, plots (a) and (d) are colored by the first and second observer's estimated state coordinates respectively, plots (b)and (e) are colored by the first and second
coordinates of the extension scheme in \cite{TalmonPNAS} and plots
(c) and (f) are colored by the first and second coordinates of the Nystr\"{o}m
extension. It is visible that the observer-based extension coordinates
are superior in describing the elevation and azimuth angles, since increasing angle values are represented by increasing observer coordinate values as seen by the distinct coloring in plots (a) and (d). In addition, the color gradient
is smoother in these plots, which shows higher correlation between
the observer coordinates and the underlying angles.


\subsection{Demonstrating the Dynamics Estimation\label{sub:Dynamics-Estimation-Example}}

In this subsection, we test the accuracy of the
dynamics parameter estimation $\Lambda$ from \eqref{eq:observer_rec}.
Here, the system is based on a toy example similar to the one presented
in Section \ref{sub:Estimation-of-the} with the following modification.
Instead of a radiating object moving on a 3D sphere, we examine the
above scheme in 2D, which depicts a radiating object moving on a 2D
circle, where the underlying 1D process is described by an azimuth
angle. We use the following dynamics equation, which has known eigenvalues,
to generate the angle: 
\[
\dot{\theta}=-\theta+\sqrt{2}\dot{w}
\]
and the 2D process is given by: 
\begin{eqnarray*}
x_{1}\left(t\right) & = & \cos\left(\theta\right)\\
x_{2}\left(t\right) & = & \sin\left(\theta\right)
\end{eqnarray*}

The eigenfunctions of the backward Fokker-Plank operator of this process
satisfy the probabilists Hermite equation \cite{risken1984fokker} which has solutions with eigenvalues $\lambda_{\ell}=-\ell,\,\ell\in\left\{ 0,1,2,...\right\} $.
The known eigenvalues provide ground truth, and as a result,
we can verify our estimation of the dynamics $\Lambda_{\ell\ell}=-\lambda_{\ell}$,
which are determined by these eigenvalues.

These modifications were performed since in the 1D setting more
complex stochastic differential equations with known solutions are
available. In addition, the effects of different components in the
proposed framework on the quality of the estimation are more noticeable
in this simpler case.

Since the accuracy of the estimation is highly dependent on the correct
calculation of the covariance matrices in the modified Mahalanobis
distance \eqref{eq:Mahalanibis Eq}, in this subsection only we simulated
short bursts at each time point for the covariance calculations. This
is performed in order to acquire a sufficient number of samples from
the same distribution at each small neighborhood and achieve a sufficiently
small error due to finite sampling.

In the setting presented in this subsection, the scale $\epsilon$
that empirically attained good performance was set to $\epsilon=0.16\zeta$
where $\zeta$ denotes the median of the pairwise distances.

\begin{figure}[t]
\begin{centering}
\includegraphics[width=0.75\columnwidth]{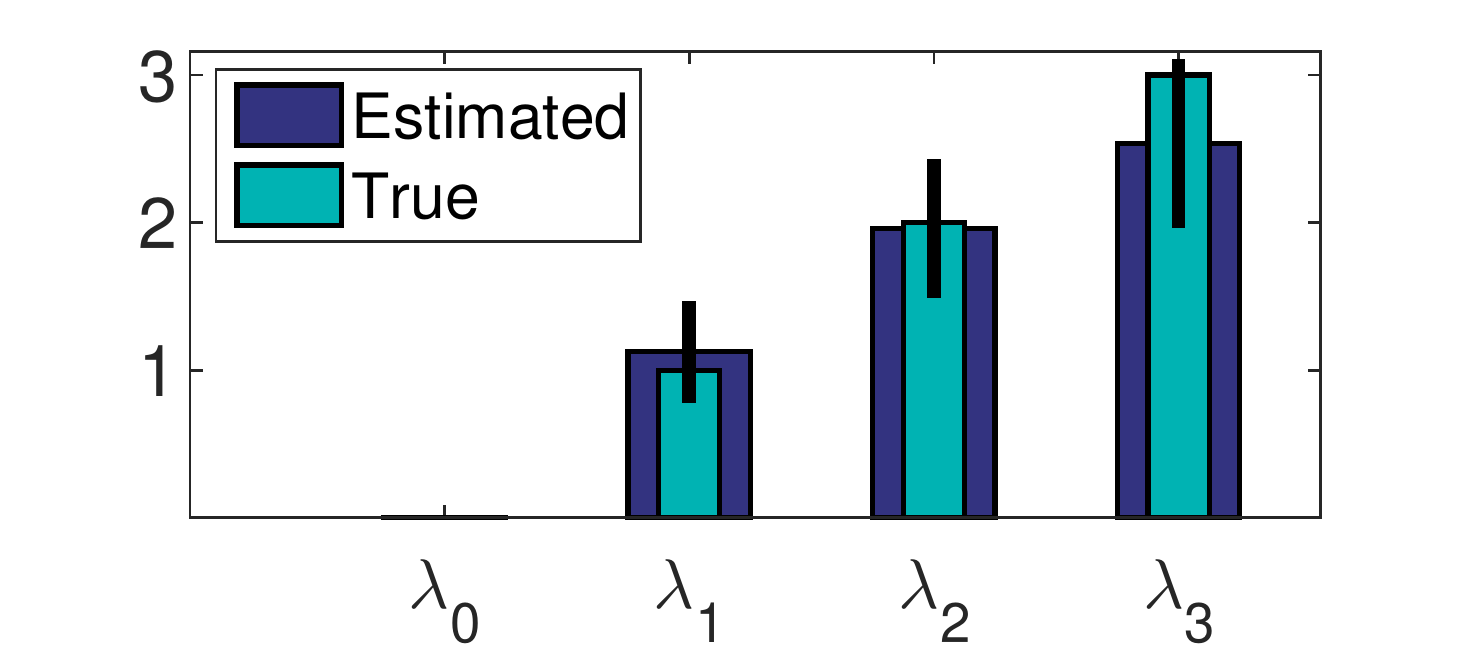} 
\par\end{centering}

\protect\protect\protect\caption{Estimation of eigenvalues $-\lambda_{\ell}$ in the 2D framework, averaged
over 50 realizations}
\label{fig:PNAS_lambda_plot} 
\end{figure}

Figure \ref{fig:PNAS_lambda_plot} presents the estimation of the
eigenvalues $-\lambda_{\ell}$ based on 1600 time frames (histograms),
each containing 60 time points which were used to construct one histogram.
The plot displays the ground truth values (green) and the average
estimates of $-\lambda_{\ell}$ (blue), over 50 realizations, for the four smallest eigenvalues (in absolute value), along with error-bars of one standard deviation
(black lines). It shows that the average dynamics estimation is
close to the true value, however, the estimation variance increases
for larger eigenvalues (in absolute value). Empirical analysis revealed that
the choice of $\epsilon$ in \eqref{eq:PairwiseAffinity} has a significant
impact on the accuracy of $\lambda_{\ell}$ estimates. Furthermore,
additive noise might also increase the estimation error, however it
mostly affects larger eigenvalues ($j=3,4,...$). Note
that this behaviour depicts one of the advantages of the presented
approach. Roughly, small eigenvalues represent signal components with
distinct ``structures'', whereas large eigenvalues tend to represent
noise.

\subsection{Music Analysis}

In this subsection we show that the proposed observer, when applied to music, reveals meaningful underlying processes describing different characteristics of the data
such as its dominant musical notes and a distinction between intervals
with different musical instruments. This is achieved without explicit
pitch tracking or modeling. For this purpose we applied Algorithm
\ref{alg:Diffusion maps discrete} and Algorithm \ref{alg:Discrete-observer-framework}
(with $\gamma=0.1$) to two songs: (i) the theme song of ``Once upon a time in the west'' performed by Yo-Yo Ma (Yo-Yo Ma plays Ennio Morricone) and
(ii) the theme song of ``The good, the bad and the ugly'' by Ennio Morricone. Both songs are sampled at 44.1 KHz and were analyzed in 15-25 second segments. In each segment
we first performed short time Fourier transform (STFT) on 23 millisecond
time frames and constructed a kernel based on the modified Mahalanobis distance
\eqref{eq:Mahalanibis Eq} between the resulting spectrograms (absolute value of the STFT). We then applied the diffusion maps framework as described in Section
\ref{sub:Manifold-Learning-and} and constructed the observer based
on the learned dynamics and linear lift function \eqref{eq:Observer_actual}.
In both songs we examined the first three coordinates of the observer's estimated state.

\begin{figure}[t]
\centering
\subfloat[]{\includegraphics[width=0.50\textwidth]{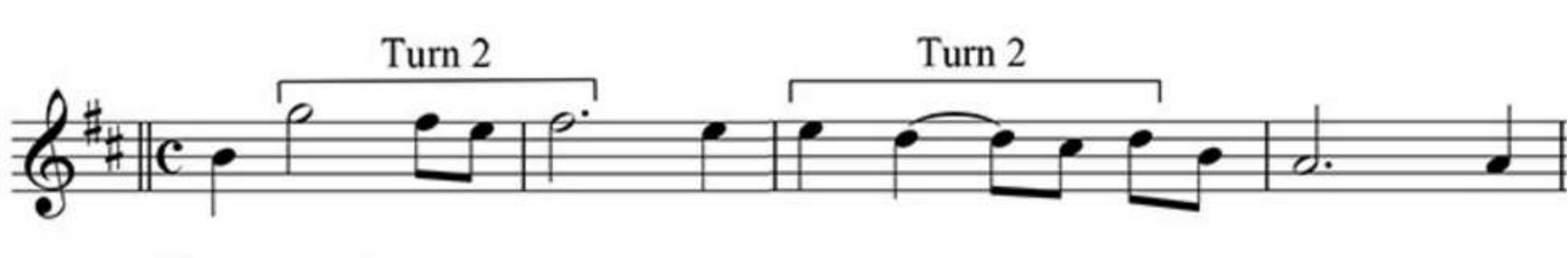}}
\vspace{-0.4cm}
\subfloat[]{\includegraphics[width=0.50\textwidth]{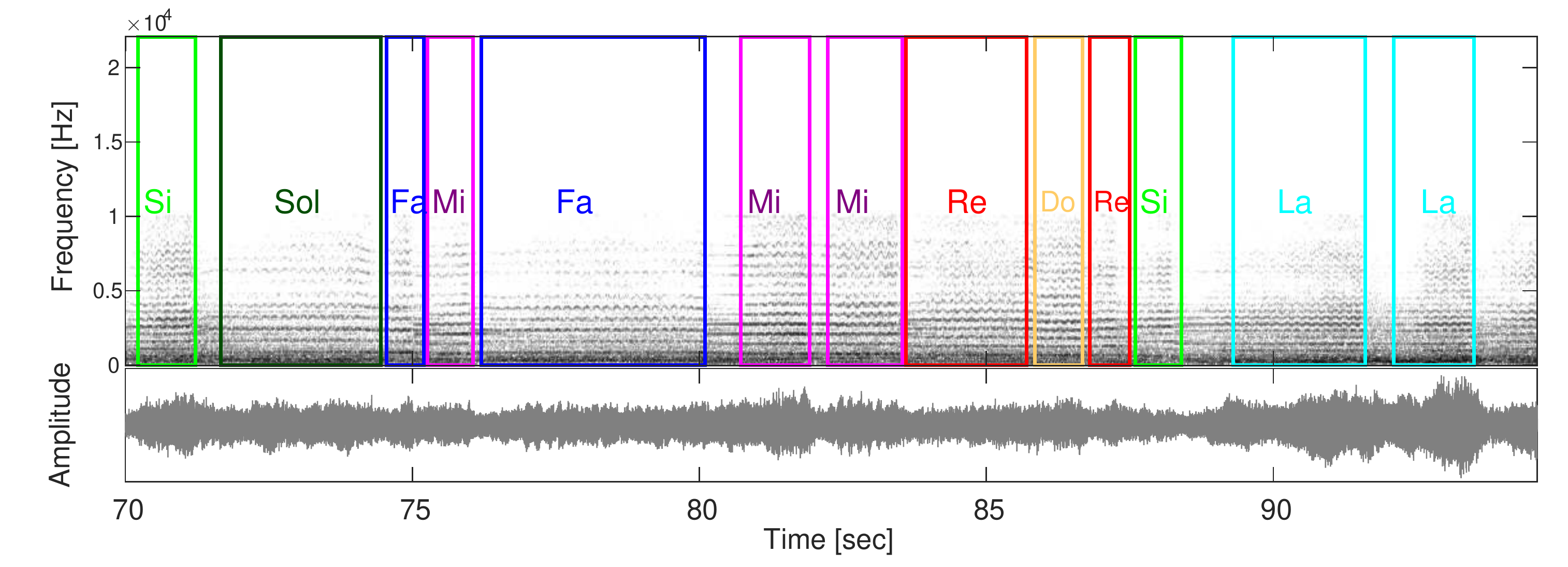} }
\vspace{-0.4cm}
\subfloat[]{\includegraphics[width=0.5\textwidth]{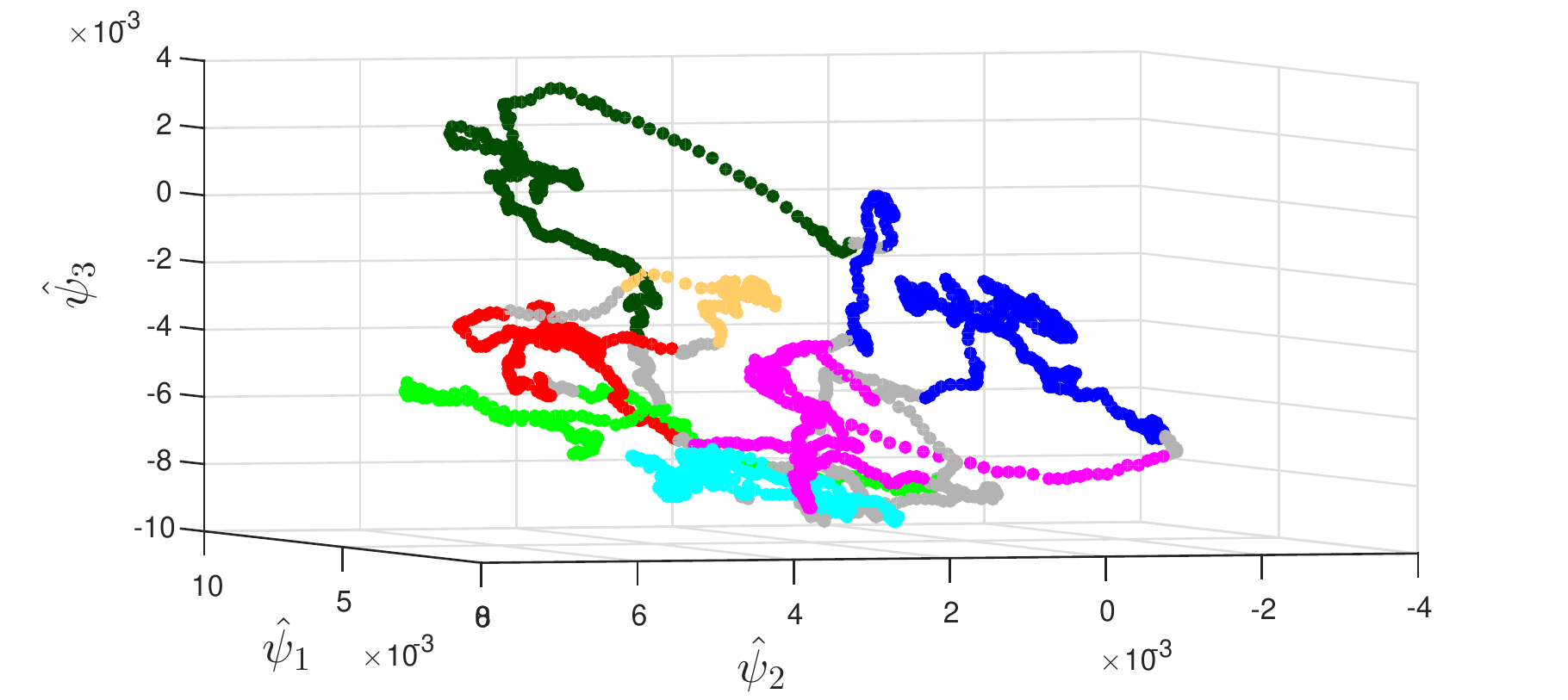} }
\vspace{-0.4cm}
\subfloat[]{\includegraphics[width=0.25\textwidth]{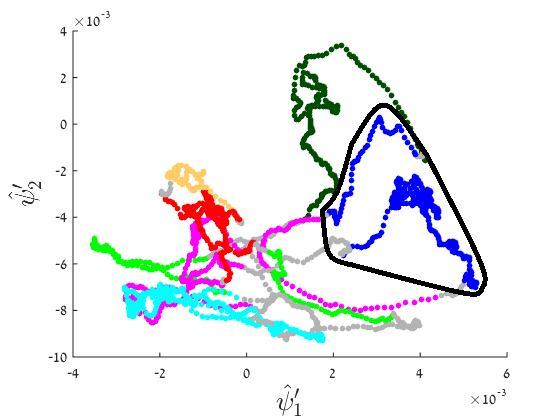}\includegraphics[width=0.25\textwidth]{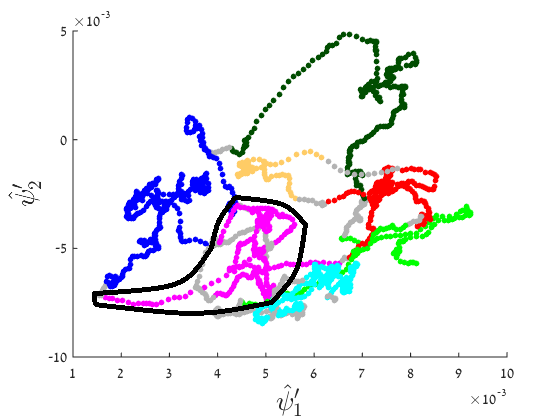} }
\vspace{-0.2cm}
\caption{Musical note identification in ``Once upon a time in the west''
theme. (a) Actual musical notes at 1:10 to 1:40 minutes. (b) Spectrogram of the music segment with marked musical notes. (c) 3D scatter plot of the first three observer coordinates. (d) Two different views of the 3D scatter plot.}
\label{fig:Music_plots_scatter} 
\end{figure}

Figure \ref{fig:Music_plots_scatter} shows a 25 second segment of
the first song (Yo-Yo Ma plays Ennio Morricone), which contains a
long segment of a single instrument, playing distinctive notes (with
a quiet background melody). The top plot of the figure (plot (a)) displays the
musical notes of the examined segment. In plot (b), the spectrogram
of the music segment is presented with marked musical notes, along with
the corresponding waveform of the music signal (below the spectrogram). Different colors
on the spectrogram represent different musical notes and correspond
to the coloring of plots (c) and (d). Plot (c) displays a 3D scatter plot of the first three observer coordinates.
Each point in this plot represents one time sample of the spectrogram
and is colored according to the marked notes. Gray points
represent transitions between notes which are unmarked on
the spectrogram. The two bottom plots (d) contain different 2D views of
the 3D scatter plot, where $\hat{\psi'}_{1},\hat{\psi'}_{2}$ represent rotated axes for better view angle. 

In this figure, the 3D scatter plot depicts that the first three coordinates
of the observer's estimated state create an embedding which separates different musical
notes to different locations in the 3D space. This is expressed as
differently colored points, representing different notes, which appear
in distinct locations in the 3D space. In addition, the two bottom
scatter plots illustrate that identical musical notes, which appear at different times,
are represented by the same color and are grouped together in the new coordinate
system created by the observer. In these bottom plots, black polygons
are marked to emphasize two such examples.

\begin{figure}[t]
\centering
\subfloat[]{\begin{centering}
\includegraphics[width=0.48\textwidth]{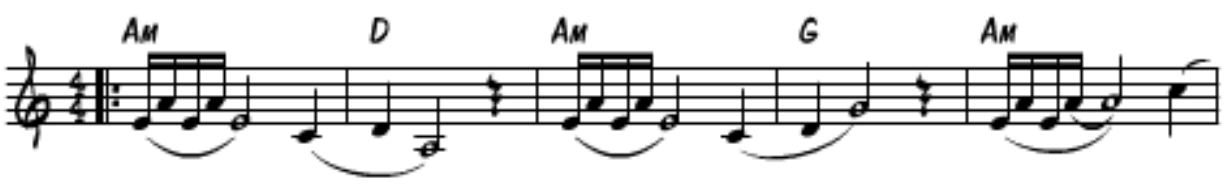}
\par\end{centering} }

\vspace{-0.2cm}
\subfloat[]{\begin{centering}
\includegraphics[width=0.5\textwidth]{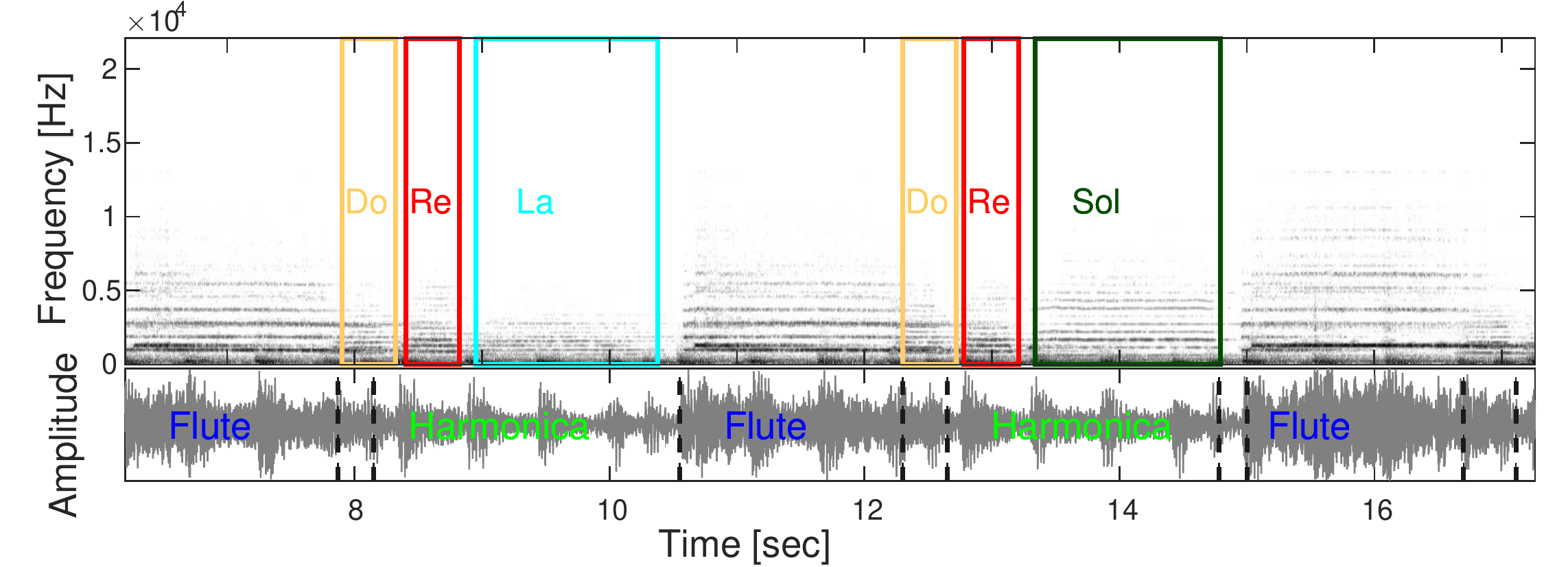}
\par\end{centering} }

\vspace{-0.2cm}
\subfloat[]{\begin{centering}
\includegraphics[width=0.5\textwidth]{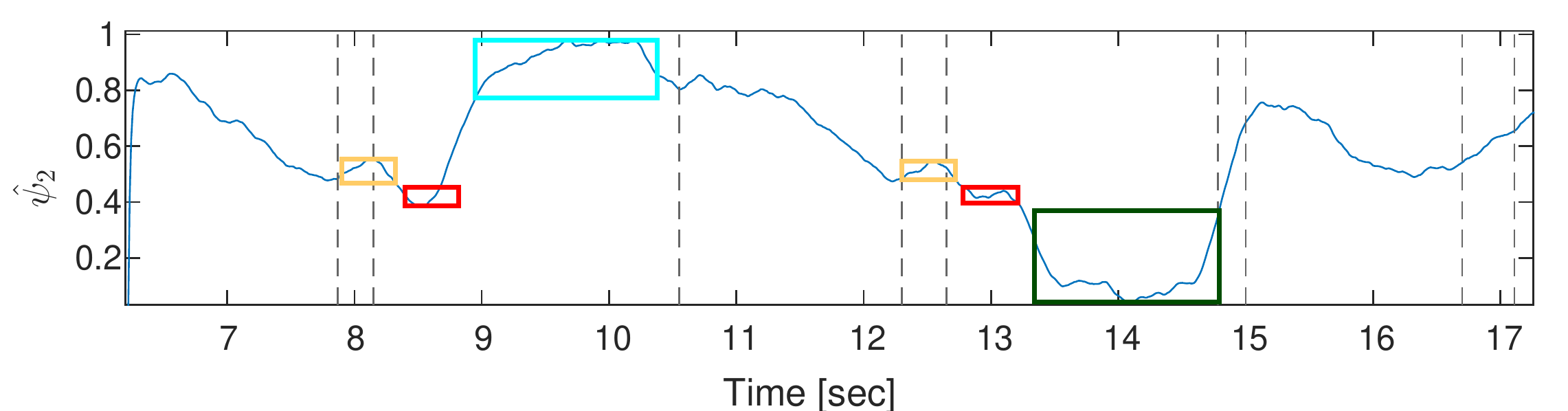}
\par\end{centering} }

\vspace{-0.2cm}
\subfloat[]{\begin{centering}
\includegraphics[width=0.5\textwidth]{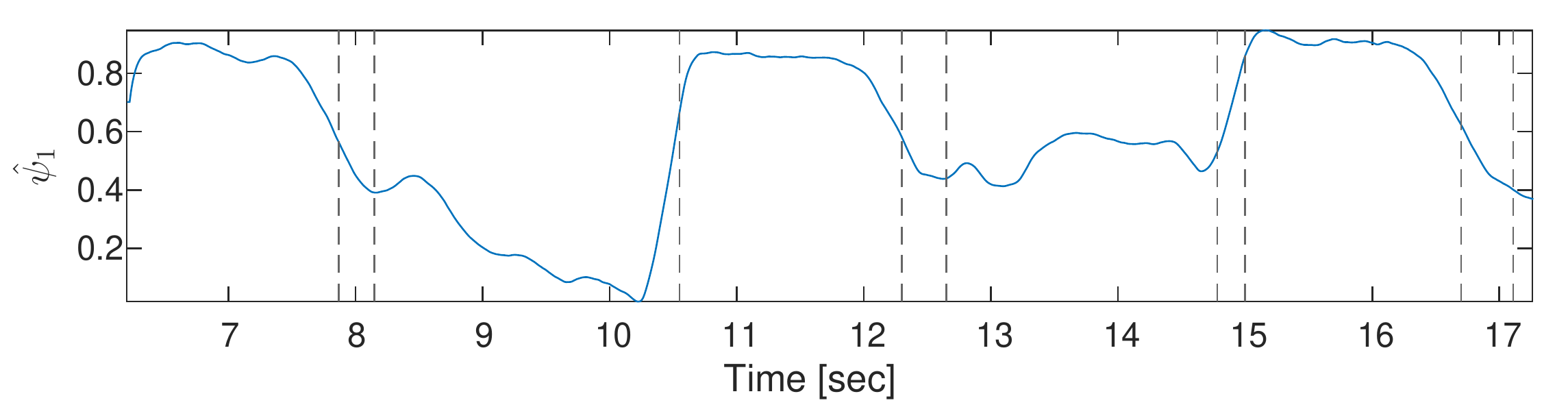}
\par\end{centering} }
\caption{Musical note and instrument identification in ``The good, the bad
and the ugly'' theme. (a) Actual musical notes at 6.2 to 17.2 seconds.
(b) Spectrogram and waveform with marked musical notes
and instruments. (c) The second observer coordinate with
marked note locations. (d) The first observer coordinate with locations of instrument transitions marked by adjacent dashed lines.}
\label{fig:Music_plots2} 
\end{figure}

Another example for musical notes identification is shown in Figure \ref{fig:Music_plots2},
which presents an 11 second segment of the second song (6.2-17.2 seconds).
This song segment contains frequent transitions between different
instruments (a flute and an harmonica). The top plot (a) in this figure
contains the musical notes of the examined segment. In plot (b) the spectrogram of the music segment is presented with marked
musical notes in the harmonica sections, along with the waveform below. On the waveform plot, the different instruments and transitions between them are marked by dashed lines
i.e., intervals where both instruments are playing. The different colors
on the spectrogram represent different musical notes and correspond
to the coloring of the plot (c). Plot (c)
presents the second observer coordinate as a function of time with
marked notes and marked transitions between the instruments.
The bottom plot (d) displays the first observer coordinate as a function
of time with marked instrument transitions.

The two bottom plots of Figure \ref{fig:Music_plots2} illustrate
that based on the observer, a good classification of both musical notes
and instruments can be attained. For example, in plot (c), the values and shape of the curves in the harmonica intervals
are consistent for similar notes at different times, e.g. Do at 7.8
seconds and at 12.1 seconds. In addition, different notes are represented
by different values, e.g. La at 8.9 seconds and Sol at 13.3 seconds.
We note that, in the flute intervals, such a distinction is not possible
since they contain rapid note changes. This leads to different time
scales in the data which are not represented properly by the embedded
coordinates (both observer coordinates and diffusion maps coordinates). 
In the example presented in Figure \ref{fig:Music_plots2}, similarly to the results shown in Figure \ref{fig:Music_plots_scatter}, the second coordinate of the observer captures the musical notes. However, since the second song includes another factor (different instruments), the first coordinate describes the instruments in this case. An example for such instrument distinction can be seen in the plot (d),
where a simple threshold value can be used to separate between the
flute intervals and the harmonica intervals.

\begin{figure}[t]
\subfloat[]{\protect\includegraphics[width=0.5\columnwidth]{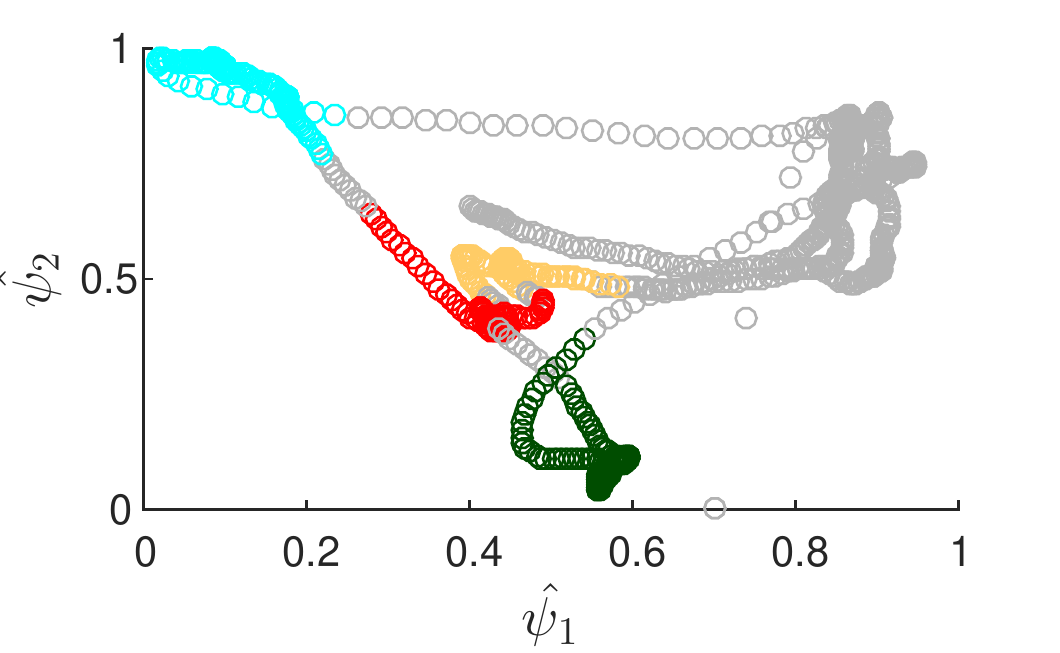} }
\subfloat[]{\protect\includegraphics[width=0.5\columnwidth]{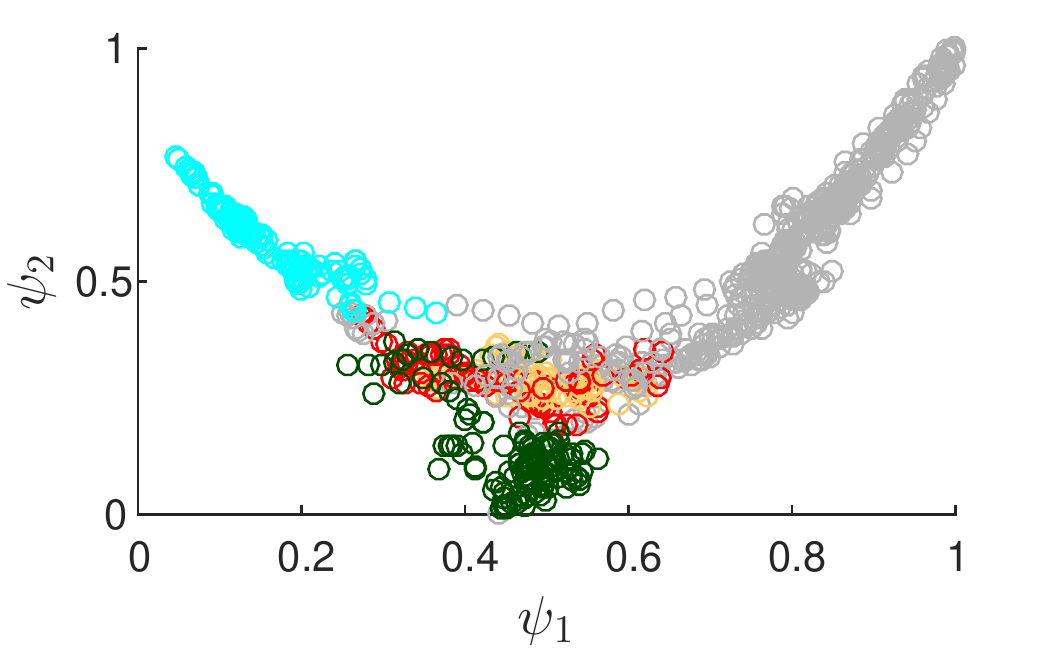} }
\vspace{-0.4cm}
\subfloat[]{\protect\includegraphics[width=0.5\columnwidth]{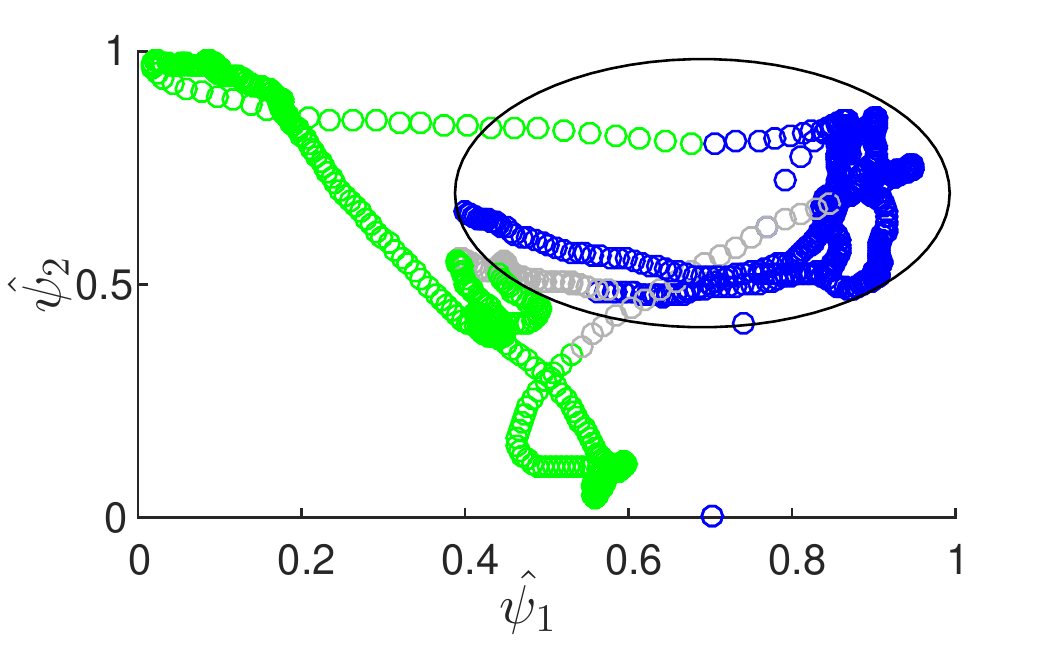} }
\subfloat[]{\protect\includegraphics[width=0.5\columnwidth]{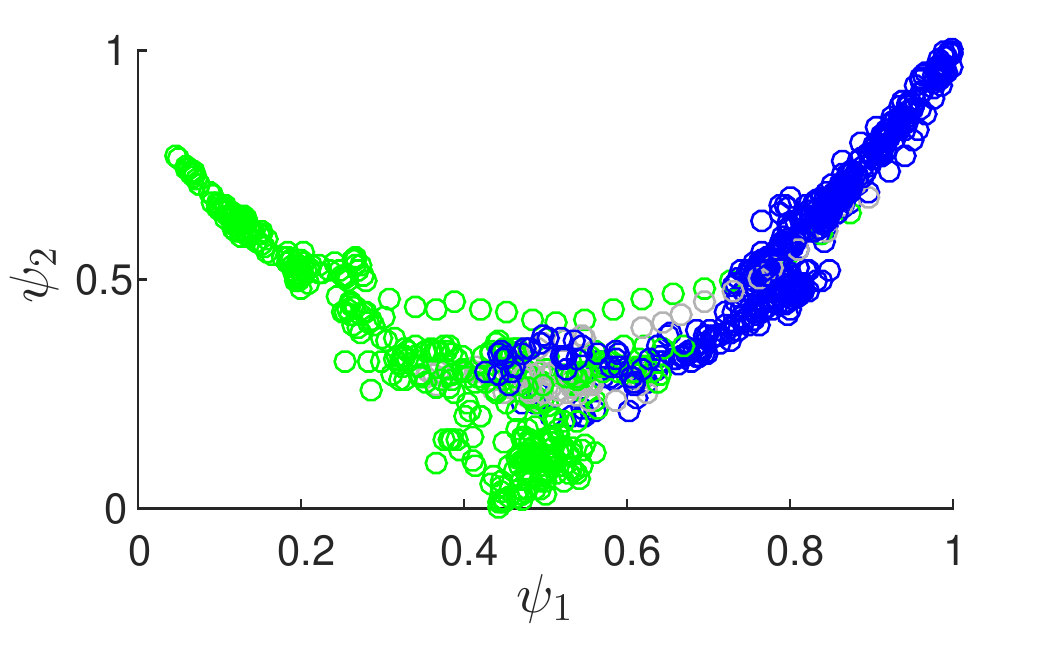} }
\vspace{-0.2cm}
\caption{\label{fig:Music_Comparison-of-observer}Comparison of observer coordinates and diffusion maps coordinates for "The good, the bad and the ugly" theme. (a,b) Scatter plot of the observer and diffusion maps coordinates respectively, colored by musical notes. (c,d) Scatter plot of the observer and diffusion maps coordinates respectively, colored by instruments: flute (blue) and harmonica (green).}
\end{figure}

Finally, in Figure \ref{fig:Music_Comparison-of-observer} we compare the coordinate systems attained by the observer and by diffusion maps for the song segment of "The good, the bad and the ugly" theme (6.2-17.2 seconds), presented in Figure \ref{fig:Music_plots2}. Figure \ref{fig:Music_Comparison-of-observer} contains two identical scatter plots of the observer coordinates in (a) and (c), $\hat{\psi}_2$ as a function of $\hat{\psi}_1$, and two identical scatter plots of the diffusion maps coordinates in (b) and (d), $\psi_2$ as a function of $\psi_1$. Each point in the plots corresponds to one time sample of the spectrogram depicted in Figure \ref{fig:Music_plots2}. The presented scatter plots differ in their color schemes. Plots (a) and (b) are colored according to the musical notes, which are marked on the spectrogram in Figure \ref{fig:Music_plots2}. In addition, here, gray points represent unmarked time samples which are either flute intervals or transitions between notes in the harmonica intervals. Plots (c) and (d) are colored according to the true instrument segmentation which is presented on the audio waveform in Figure \ref{fig:Music_plots2}. This color scheme marks flute intervals in blue, harmonica intervals in green and transitions between them, in which both instruments are playing, in gray. In the displayed plots, the diffusion maps coordinates and the observer coordinates were scaled for comparison only.
Figure \ref{fig:Music_Comparison-of-observer} depicts that the coordinate system attained by the observer represents the data more accurately than the one attained by diffusion maps. This is visible in plots (a) and (b) where the coloring of the notes in plot (a) (observer coordinates) implies a better distinction compared to plot (b). Furthermore, the two separate segments containing Do and Re in the spectrogram in Figure \ref{fig:Music_plots2} are correctly grouped together in plot (a). In addition, plots (c) and (d) depict that the observer coordinates perform better in the classification of different music instruments as the blue and green colored points are highly separable.

\section{Conclusions}\label{sec:Conclusions}
In this work we presented a purely data driven framework, in which an intrinsic representation is derived for highly nonlinear noisy systems. 
The framework consists of a combination between a manifold learning technique, which is used to reveal the hidden parameters of the system, and a linear observer, that incorporates  dynamics and attains the intrinsic representation.
We showed that our method indeed reveals intrinsic hidden features in data. In particular, when applied to music, it discovers both a segmentation to different musical notes as well as an indication of various musical instruments.

In future work we plan to address two main aspects. First, we plan to implement a Kalman filtering scheme that will provide an optimal adaptive choice of the gain parameter $\kappa$. Such a scheme will have many advantages, for example, it will circumvent the need to empirically tune the weighting parameter $\gamma$.
The second issue we will address concerns the dynamics estimation. As presented in the experimental results section, the estimation of the dynamics based on diffusion maps might be influenced by measurement noise. In addition, in our work, we focused on the linear component of the diffusion maps dynamics and compensated for the stochastic component as part of the measurement correspondence element. This can be sufficient in systems where the linear dynamics component is dominant. When the stochastic component is dominant, we plan to extend our observer framework by adding an adaptive linear dynamics parameter which will account for these errors.

\section*{Acknowledgements}
The authors would like to thank Professor Ronald Coifman for fruitful discussions. 

\appendices{}

\section{Eigenfunction Dynamics\label{sub:Appendix Eigenfunction-Dynamics}}
In this appendix we show the derivation of the stochastic differential equation for the eigenfunctions of the backward Fokker-Planck operator presented in Section \ref{sub:Revealing-the-dynamics}. Recall the state equation presented in Section \ref{sec:Problem-Formulation}:
\begin{equation}
\dot{\boldsymbol{\theta}}(t)=-\nabla U\left(\boldsymbol{\theta}\left(t\right)\right)+\sqrt{\frac{2}{\beta}}\dot{\boldsymbol{\omega}}\left(t\right)
\end{equation}
Since the eigenfunctions of the Fokker-Planck operator which describes this system are smooth functions of the underlying state, based on It\^o calculus they also evolve according to a stochastic differential equation of the form
\begin{equation}
\dot{\psi}_{\ell}\left(\boldsymbol{\theta}\left(t\right)\right)=\mu_{\ell}\left(\boldsymbol{\theta}\left(t\right)\right)+\sigma_{\ell}\left(\boldsymbol{\theta}\left(t\right)\right)\dot{\omega}_{\ell}\left(t\right)
\end{equation}
where $\mu_{\ell}\left(\boldsymbol{\theta}\left(t\right)\right)$, $\sigma_{\ell}\left(\boldsymbol{\theta}\left(t\right)\right)$ are the drift and diffusion coefficients respectively and $\omega_{\ell}\left(t\right)$ is Brownian motion.
According to It\^o's lemma, assuming that the Brownian motions of different coordinates of $\boldsymbol{\theta}\left(t\right)$ are independent, the drift and diffusion parameters are given by
\begin{eqnarray}
\mu_{\ell}\left(\boldsymbol{\theta}\left(t\right)\right) & = & \frac{\partial \psi_{\ell}}{\partial t}+\frac{1}{\beta}\triangle_{\theta}\psi_{\ell}-\nabla_{\theta}U\cdot\nabla_{\theta}\psi_{\ell}\label{eq:Eigen_drift}\\
\sigma_{\ell}\left(\boldsymbol{\theta}\left(t\right)\right) & = & \sqrt{\frac{2}{\beta}}\left\Vert\nabla_{\theta}\psi_{\ell}\right\Vert
\end{eqnarray}
The partial derivative of $\psi_{\ell}$ in time is zero, since the eigenfunctions depend only on the state parameter and describe the system in steady state. Therefore, we are left with the two terms in \eqref{eq:Eigen_drift} which are exactly the left-hand side of the backward Fokker-Planck operator \eqref{eq:Fokker_Planck}. Since $\psi_{\ell}$ are the eigenfunctions of this operator with corresponding eigenvalues $-\lambda_{\ell}$, the resulting drift is described by $\mu_{\ell}\left(\boldsymbol{\theta}\left(t\right)\right)=-\lambda_{\ell}\psi_{\ell}\left(\boldsymbol{\theta}\left(t\right)\right)$. 

Therefore, the stochastic differential equation, describing the dynamics of the eigenfunctions is
\begin{equation}
\dot{\psi}_{\ell}\left(\boldsymbol{\theta}\left(t\right)\right)=-\lambda_{\ell}\psi_{\ell}\left(\boldsymbol{\theta}\left(t\right)\right)+\sqrt{\frac{2}{\beta}}\left\Vert\nabla_{\theta}\psi_{\ell}\left(\boldsymbol{\theta}\left(t\right)\right)\right\Vert
\end{equation}

\section{Mahalanobis Distance\label{sub:Appendix_Mahalanobis-Distance}}
We present the derivation of the modified Mahalanobis distance \cite{Singer2007} described in Section \ref{sub:Non-linear-Measurement-mapping}, which approximates the Euclidean distances of $\boldsymbol{\theta}\left(t\right)$ from the measurements $\boldsymbol{z}\left(t\right)$. 

For the noiseless case, consider the Taylor expansion at $\tau$ of the measurement function \eqref{eq:MeasurementEq}
\begin{eqnarray}
\boldsymbol{z}\left(t\right) & = & h\left(\boldsymbol{\theta}\left(\tau\right)\right) + J_h\left(\boldsymbol{\theta}\left(\tau\right)\right)\left(\boldsymbol{\theta}\left(t\right)-\boldsymbol{\theta}\left(\tau\right)\right)\nonumber\\
& & +O\left(\left\Vert\boldsymbol{\theta}\left(t\right)-\boldsymbol{\theta}\left(\tau\right)\right\Vert^2\right)
\end{eqnarray}
where $J_h\left(\boldsymbol{\theta}\left(\tau\right)\right)$ is the Jacobian matrix of $h\left(\boldsymbol{\theta}\left(\tau\right)\right)$.
Applying the squared Euclidean norm and taking the inverse of the Jacobian we get
\begin{eqnarray}
\left\Vert\boldsymbol{\theta}\left(t\right)-\boldsymbol{\theta}\left(\tau\right)\right\Vert^2 & = & \left\Vert J_h^{-1}\left(\boldsymbol{z}\left(\tau\right)\right)\left(\boldsymbol{z}\left(t\right)-\boldsymbol{z}\left(\tau\right)\right)\right\Vert^2 \nonumber\\
& & +O\left(\left\Vert\boldsymbol{\theta}\left(t\right)-\boldsymbol{\theta}\left(\tau\right)\right\Vert^4\right)
\end{eqnarray}
Similarly, we can derive the Taylor expansion at $t$ and average the two resulting terms:
\begin{eqnarray}
\left\Vert\boldsymbol{\theta}\left(t\right)-\boldsymbol{\theta}\left(\tau\right)\right\Vert^2 & = & \frac{1}{2}\left\Vert J_h^{-1}\left(\boldsymbol{z}\left(\tau\right)\right)\left(\boldsymbol{z}\left(t\right)-\boldsymbol{z}\left(\tau\right)\right)\right\Vert^2\nonumber\\
& + & \frac{1}{2}\left\Vert J_h^{-1}\left(\boldsymbol{z}\left(t\right)\right)\left(\boldsymbol{z}\left(t\right)-\boldsymbol{z}\left(\tau\right)\right)\right\Vert^2\nonumber\\
& + & O\left(\left\Vert\boldsymbol{\theta}\left(t\right)-\boldsymbol{\theta}\left(\tau\right)\right\Vert^4\right)\label{eq:First_Step_Mahalanobis}
\end{eqnarray}
The expression in \eqref{eq:First_Step_Mahalanobis} can be written as
\begin{align}
\left\Vert\boldsymbol{\theta}\left(t\right)-\boldsymbol{\theta}\left(\tau\right)\right\Vert^2 \nonumber
& =\frac{1}{2}\left(\boldsymbol{z}\left(t\right)-\boldsymbol{z}\left(\tau\right)\right)\cdot M\left(t,\tau\right)\cdot\left(\boldsymbol{z}\left(t\right)-\boldsymbol{z}\left(\tau\right)\right)^T\nonumber\\
& +O\left(\left\Vert\boldsymbol{\theta}\left(t\right)-\boldsymbol{\theta}\left(\tau\right)\right\Vert^4\right)\label{eq:AlmostMahalanobis}
\end{align}
where $M\left(t,\tau\right)=\left(J_{h}J_{h}^T\right)^{-1}\left(\boldsymbol{z}\left(t\right)\right)+\left(J_{h}J_{h}^T\right)^{-1}\left(\boldsymbol{z}\left(\tau\right)\right)$.

Lastly, to obtain the modified Mahalanobis distance in \cite{Singer2007}, we show that $\left(J_{h}J^{T}_{h}\right)\left(\boldsymbol{z}\left(t\right)\right)$ is in fact the covariance matrix of the measurements at time $t$, $C\left(\boldsymbol{z}\left(t\right)\right)$. 

Since the state of the system $\boldsymbol{\theta}\left(t\right)$ satisfies the stochastic differential equation in \eqref{eq:StochEq} and $\boldsymbol{z}\left(t\right)=h\left(\boldsymbol{\theta}\left(t\right)\right)$, based on It\^o's lemma the measurements satisfy
\begin{eqnarray}
dz_j\left(t\right)=\sum_{i=1}^{d}\left(\frac{1}{2}\left(b_{i}\right)^{2}\frac{\partial^2 h_j}{\partial\theta_i^2}+a_{i}\frac{\partial h_j}{\partial \theta_i}\right)dt+\sum_{i=1}^{d}b_{i}\frac{\partial h_j}{\partial\theta_i}d\omega_i\nonumber
\end{eqnarray}
where $j=1,...,n$, $a_i$ and $b_i$ are the drift function and diffusion coefficient of coordinate $\theta_i\left(t\right)$, as presented in \eqref{eq:StochEq} and $\omega_i$ is Brownian motion.
Assuming that the Brownian motions of different coordinates $\theta_i\left(t\right)$ are independent, the covariance matrix of the measurements is given by
\begin{eqnarray}
C_{jk}\left(\boldsymbol{z}\left(t\right)\right) & = & \sum_{i=1}^{d}\left(b_{i}\right)^{2}\frac{\partial h_j}{\partial\theta_i}\frac{\partial h_k}{\partial\theta_i},\hspace{0.5em} j,k = 1,...,n
\end{eqnarray}
Note that the diffusion coefficients in our setting are constant, therefore, we can first apply a scaling transformation to eliminate $b_i$ as described in \cite{Singer2007}.
Finally, after this scaling, the covariance can be written using the Jacobian matrix $J_h$:
\begin{equation}
C\left(\boldsymbol{z}\left(t\right)\right)=J_{h}J_{h}^T\left(\boldsymbol{z}\left(t\right)\right)\label{eq:CovarianceJ}
\end{equation}
By placing \eqref{eq:CovarianceJ} in \eqref{eq:AlmostMahalanobis} we attain the desired form of the modified Mahalanobis distance.

\bibliographystyle{IEEEbib}
\bibliography{papers}

\end{document}